\documentclass[twocolumn]{aastex61}
\usepackage{amsmath}
\usepackage{listings}
\usepackage[toc,page]{appendix}

\begin{document}

\title{WOMBAT: A Scalable and High Performance Astrophysical MHD Code}

\author{P. J. Mendygral}
\affiliation{Cray Inc., St. Paul, MN 55101}
\affiliation{School of Physics and Astronomy, University of Minnesota, Minneapolis, MN 55455}
\email{pjm@cray.com}

\author{N. Radcliffe}
\affiliation{Cray Inc., St. Paul, MN 55101}
\email{nradclif@cray.com}

\author{K. Kandalla}
\affiliation{Cray Inc., St. Paul, MN 55101}
\email{kkandalla@cray.com}

\author{D. Porter}
\affiliation{Minnesota Supercomputing Institute for Advanced Computational Research}
\email{dhp@umn.edu}

\author{B. J. O'Neill}
\affiliation{School of Physics and Astronomy, University of Minnesota, Minneapolis, MN 55455}
\affiliation{Minnesota Supercomputing Institute for Advanced Computational Research}
\email{oneill@astro.umn.edu}

\author{C. Nolting}
\affiliation{School of Physics and Astronomy, University of Minnesota, Minneapolis, MN 55455}
\affiliation{Minnesota Supercomputing Institute for Advanced Computational Research}
\email{nolt0040@umn.edu}

\author{P. Edmon}
\affiliation{Institute for Theory and Computation, Center for Astrophysics, Harvard University, Cambridge, MA 02138}
\email{pedmon@cfa.harvard.edu}

\author{J. M.F. Donnert}
\affiliation{School of Physics and Astronomy, University of Minnesota, Minneapolis, MN 55455}
\affiliation{INAF-Istituto di Radioastronomia, via. P.Gobetti 101, I-40129 Bologna Italy}
\affiliation{ERC Marie Curie Fellow}
\email{donnert@ira.inaf.it}

\author{T. W. Jones}
\affiliation{School of Physics and Astronomy, University of Minnesota, Minneapolis, MN 55455}
\affiliation{Minnesota Supercomputing Institute for Advanced Computational Research}
\email{twj@umn.edu}


\begin{abstract}
We present a new code for astrophysical magneto-hydrodynamics specifically designed and optimized for high performance and scaling on modern and future supercomputers.  We describe a novel hybrid OpenMP/MPI programming model that emerged from a collaboration between Cray, Inc. and the University of Minnesota.  This design utilizes MPI-RMA optimized for thread scaling, which allows the code to run extremely efficiently at very high thread counts ideal for the latest generation of the multi-core and many-core architectures.  Such performance characteristics are needed in the era of ``exascale'' computing.  We describe and demonstrate our high-performance design in detail with the intent that it may be used as a model for other, future astrophysical codes intended for applications demanding exceptional performance.

\end{abstract}



\section{Introduction} \label{introduction}

Magneto-hydrodynamic (MHD) simulations allow us to study the dynamics of highly conducting astrophysical fluids since many astrophysical fluids are highly conductive ionized plasmas. MHD modeling then allows us to incorporate essential consequences of magnetic fields.  Even ``weak'' magnetic fields, whose Maxwell stresses are subdominant to inertial and to thermal pressure stresses, can have major impact on the development of turbulence and its dissipation on small scales, on momentum transport, angular momentum and energy, and on thermal conduction. If the simulations include, in addition to MHD, the transport of high energy, non-thermal ``cosmic ray'' particle populations, the simulations can model emission processes involving the cosmic ray interactions with the bulk fluid and its magnetic field. These include $\gamma$-ray by-products of cosmic ray proton interactions with the bulk fluid and radio to X-ray emissions from cosmic ray electrons, including synchrotron radiation.\par
Since magnetic field properties often derive from the details of the fluid dynamics over a wide range of scales of interest, it is essential for simulations to capture the dynamics with high fidelity over this full range of scales. This is generally a very intensive and challenging computational task that, despite much progress in coding and vast improvements in computing infrastructure, has often remained beyond current capabilities. That challenge is the motivation for our efforts described here to develop an MHD code environment that can effectively utilize and adapt to the coming generations of computational infrastructure to allow solutions to these pressing astrophysical problems.\par
Numerous codes exist for both general purpose and specific use astrophysical fluid simulations.  Some examples are GADGET \citep{2005MNRAS.364.1105S, 2009MNRAS.398.1678D}, NDSPMHD \citep{2012JCoPh.231..759P}, AREPO \citep{2010MNRAS.401..791S, 2016MNRAS.463..477M}, ENZO \citep{2014ApJS..211...19B}, ATHENA \citep{2008ApJS..178..137S, 2008JCoPh.227.4123G}, RAMSES \citep{2002A&A...385..337T, 2006A&A...457..371F}, CHARM \citep{2011ApJS..195....5M}, PLUTO \citep{2007ApJS..170..228M, 2012ApJS..198....7M}, CASTRO \citep{2010ApJ...715.1221A}, and FLASH \citep{2000ApJS..131..273F, 2008ASPC..385..145D}.  Codes like these have been developed over many years and often have features for adding the effects of gravity, cosmic-ray transport, non-ideal MHD, cosmic expansion, and non-adiabatic energy gains and losses, including radiative and conductive cooling and heating.  ``Exascale'' is next next major step in the evolution of high performance computing (HPC), with systems capable of performing 10$^{18}$ floating point operations per second distributed across many levels of parallelism.  Preparing applications for exascale requires a substantial investment in code re-design and optimization, to enable the community to leverage the capability of new architectures and make new scientific breakthroughs.  \citet{2016arXiv161008833D} recently presented a survey of the challenges and potential approaches to modernizing some of the most popular community codes.\par
The latest multi-core and many-core processors (CPUs), such as Intel Xeon and Intel Xeon Phi, feature increasing core counts per processor with decreasing clock speed along with increasing single instruction-multiple data (SIMD) vector lengths. Hence, cache blocking and vectorization are critical to obtaining good performance from modern processors. But the increasing core counts also put pressure on the traditional MPI-only (Message Passing Interface) parallelization models.  Memory consumption from a large number of independent MPI processes on a node may become prohibitive. For MHD simulations that develop substantial load imbalance, possibly through the inclusion of N-body dynamics or multi-level mesh refinement, balancing work between MPI ranks is critically important.  However, the process of balancing work between MPI ranks carries potentially significant overhead.  This overhead is the combination of the cost of moving grid data between MPI ranks and communication of the change in decomposition to some or all MPI ranks.  There are several established strategies for reducing the overhead, including decomposition meta-data replication, but these techniques come at the cost of memory and complexity \citep{2016arXiv161008833D}.  Programming models that allow for load balancing with less explicit communication are greatly needed.  \par
One attractive approach is the hybrid OpenMP/MPI model, as discussed in \citep{2014ApJS..211...19B}. It allows MPI ranks to hold larger portions of the world grid. In the context of mesh refinement, added work due to refinement at any single location is a lower fraction of a rank's total load.  For many calculations it could also result in a more symmetric load across MPI ranks if refinement needs are not confined to a single region. Work within an MPI rank can be load balanced among threads with any form of dynamic work scheduling.  Finally, on-node imbalances due to contention of shared resources, such as cache or bandwidth, also can be mitigated with attention to thread scheduling. However, typical parallel loop-based OpenMP designs have shown too little scope (amount of code effectively threaded) to scale effectively to high thread count. \par
Modern HPC interconnects often feature low latency/high bandwidth messaging with network-offloading, which enables overlap of computation with communication.  MPI-RMA (Remote Memory Access) is a feature added to the MPI standard in order to expose these capabilities to the user.  It should be possible for an application to drive communication near hardware limits with a highly efficient MPI-RMA implementation. However, MPI libraries need high performance \textbf{\mbox{MPI\_THREAD\_MULTIPLE}} implementations for the hybrid OpenMP/MPI model to include communication parallelization. \par
In this paper we present an application design study for a new grid-based MHD code called WOMBAT\footnote{WOMBAT is available by request or by visiting http://www.astro.umn.edu/groups/compastro}.  The goal of this project is to address the optimization opportunities discussed above through a co-design process.  In pursuit of this goal, we seek a base design well suited for uni-grid simulations yet formulated for complex conditions requiring load balancing.  For the purpose of this paper, we review the base design for MHD uniform meshes only.  WOMBAT development is a collaboration between Cray Inc. Programming Environments and the University of Minnesota.  Through this collaboration we developed a design strategy (see \S \ref{design}) that adapts to architectures (CPU and interconnect) using language, OpenMP and MPI best practices.  We also identified bottlenecks and optimizations for MPI (Cray MPICH) resulting in significant performance improvements. Section \ref{performance} is a performance review of WOMBAT on three architectures that can be used as a model for assessing the quality of any similar implementation.  We discuss specific implementation details in \S \ref{design_details}.  Our design strategy is applicable to many other codes and serves as a potential path forward for exascale application readiness.\par
In what follows ``KNL'' designates the Intel\textsuperscript{\textregistered} Xeon\textsuperscript{\textregistered} Phi many-core processor (Knights Landing), ``Broadwell'' a recent Intel\textsuperscript{\textregistered} Xeon\textsuperscript{\textregistered} multi-core processor (Broadwell) and ``Interlagos'' the AMD Opteron\textsuperscript{TM} multi-core processor (Interlagos). In all figures, these processors are shown as red, green and blue, respectively.


\section{Scalable Design Strategy} \label{design}

\begin{deluxetable*}{ccc}
\centering
\tablecolumns{3}
\tabletypesize{\scriptsize}
\tablecaption{HPC architecture constraints and the optimization techniques and features developed in WOMBAT to address them. \label{tab:overview}}
\tablehead{
    \colhead{Hardware Constraint}
  & \colhead{Optimization Approach}
  & \colhead{WOMBAT Design}
}
\startdata
high FLOP/Byte ratio       &  cache blocking                            &  \textbf{Patch} \\
slow scalar + wide vectors &  vectorization                             &  \textbf{Fortran} + vectorization best practices    \\
many cores                 &  thread scalability                        &  SPMD OpenMP + new Cray MPICH + \textbf{Patch}    \\
distributed architecture   &  RDMA + computation/communication overlap  &  AIO + MPI-RMA + new Cray MPICH + \textbf{Patch} \\
\enddata
\end{deluxetable*}

The key design characteristic of WOMBAT is to subdivide the problem into completely independent pieces of the world grid that include their own boundary zones and necessary meta-data for updating from one time step to the next.  We refer to these independent pieces as ``Patches.'' This design naturally accommodates any numerical method with local or semi-local communication needs. \par
The concept is similar to data management strategies in other many MHD codes \citep[see][]{2016arXiv161008833D}, but our design takes a unique approach to processing and scheduling the computation and communication of Patches.  A Patch is a unit of work that a thread within a WOMBAT MPI process independently operates on.  No assumptions are made on the number of Patches relative to the number of threads since our design adapts to this ratio.  Patch boundaries also define units of communication work done with either local (intra-process) or remote (inter-process) copies.  The number of zones in each dimension of a Patch and  the number of them in each dimension on a rank (and Domain, see \S \ref{decomposition}) are input parameters. This allows us to tune them for performance on a given architecture (see \S \ref{patch_optimization}). \par
WOMBAT is written in \textbf{Fortran 2008}.  \textbf{Fortran} semantics make it easy for modern compilers to identify and apply optimizations, such as vectorization, as long as developers follow simple rules (see \S \ref{vectorization}).  Code modularity, organization and maintainability benefit from the object-oriented features available in \textbf{Fortran 2008}. However, overuse of classes can lead to a loss of optimization opportunities for a compiler.  Thus any performance critical section of WOMBAT is basic \textbf{Fortran} code working on arrays. The code is constructed from three main categories of classes: data managers, engines and solvers.  Data managers do memory management and supporting functions.  Solvers accept data managers as arguments and update their arrays following whatever numerical methods they employ.  Engines orchestrate parallelism and the book-keeping and communication requirements for handing data managers to solvers. \par
Table \ref{tab:overview} shows the hardware constraints on the current and next generation of HPC systems, alongside the techniques and optimizations we include in our design strategy to meet these constraints.

\subsection{Domain Decomposition} \label{decomposition}

\begin{figure}
\centering
\includegraphics[height=.35\textheight]{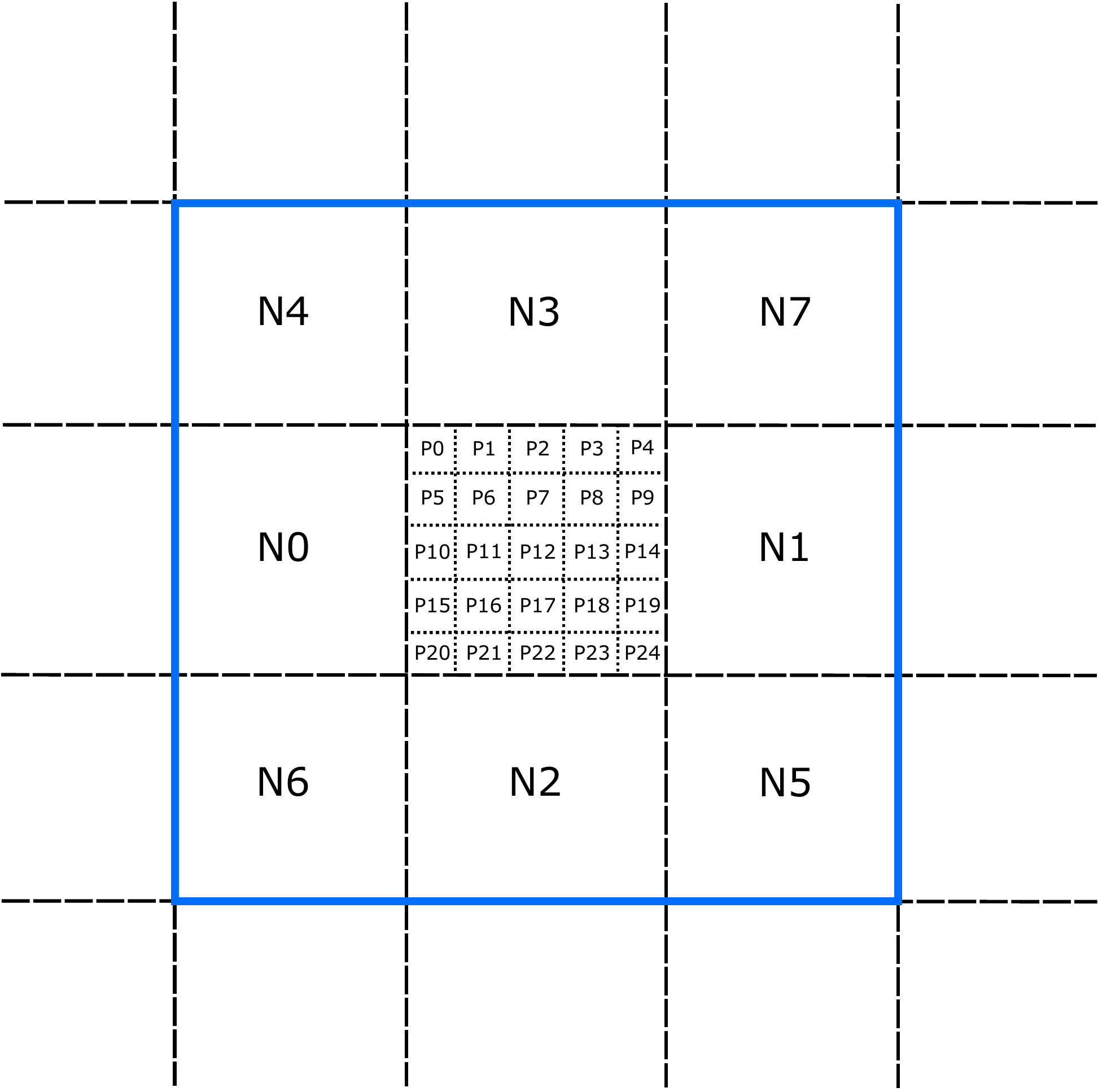}
\caption{Sample 2d decomposition.  Regions labeled N[0-7] are the Domain of each MPI neighbor rank for the MPI rank in the center.  Regions labeled P[0-24] are 
25 Patches inside each MPI rank's Domain.} \label{fig:decomp}
\end{figure}

To construct the Patches, the world grid is decomposed into N equal size sub-volumes (Domains), each assigned to an MPI rank.  An MPI rank's sub-volume is further decomposed into Patches containing an equal number of zones.  Patches in a Domain communicate boundary data with one another and with Patches on neighboring ranks. Figure \ref{fig:decomp} shows a sample 2d configuration of the Patch-Domain hierarchy across an arbitrary number of MPI ranks. The figure is centered on a single MPI rank's Domain.  That rank has eight neighbors labeled N0 through N7 each with their own Domain.  Inside every MPI rank's Domain is a 5x5 grid of Patches, labeled P0 through P24 for the central Domain. \par
We implement the Patch-Domain design as \textbf{Fortran} classes. The Domain class is responsible for tracking the MPI rank and local or remote Patches that share boundaries with it.  This design can be extended to nested block static or adaptive mesh refinement (SMR or AMR).  In support of that and generic load balancing between MPI ranks, Patches inside a Domain are allowed to become active or inactive, meaning member data structures can be allocated and updated in time, deallocated and not included in updates.  Information about the MPI rank(s) and remote Patches sharing a boundary with a Patch can be modified over time, allowing Patches to be moved between ranks with minimal bookkeeping and communication.  In particular, we do not store global data structures for tracking decomposition.

\subsection{Optimization and Multi-level Parallelization Strategy} \label{parallelism}

Three levels of parallel optimization are common to HPC systems: SIMD vectorization, intra-process threading (using OpenMP), and inter-process communication (using MPI). Cache blocking is an additional optimization that addresses memory topology on these systems.

\subsubsection{Cache Blocking} \label{cache_blocking}

The Patch design naturally results in cache blocking, which directly addresses bandwidth limitations. The most popular CPUs used in HPC today have FLOP/s to Byte/s ratios (ratio of floating point performance to memory bandwidth\footnote{For example, a 1.4 GHz Intel Xeon Phi processor is theoretically able to achieve $\simeq$ 3 TFLOP/s and has a memory bandwidth of $\simeq$ 450 GB/s to MCDRAM (90 GB/s to DDR).}) of $\simeq$ 10, hence algorithms with similar computational intensities run most efficiently. However, computational intensities that high are difficult to achieve with stencil based numerical methods for solving MHD.  Processor caches can mitigate this issue when used effectively and stencil methods provide good opportunity for reuse of loaded values between operations. Hence, good performance requires cache blocking techniques on all key loops. However, explicitly programmed cache blocking can be cumbersome, because \emph{all performance-critical} nested loops over problem dimensions must be expanded into higher dimensional loops with tunable blocking parameters. \par
Since all solvers in WOMBAT operate on a single Patch, their computationally intensive loops are all roughly the size of a Patch. The best Patch size that fits into a level of cache  inherently gets reuse out of cache (typically the best size fits into level 3 (L3) cache but not entirely into level 2 (L2), see \S \ref{patch_optimization}).

\subsubsection{SIMD Vectorization} \label{vectorization}

Operations on a Patch consist of floating point and data motion intensive loops. These loops are written to be good SIMD vectorization candidates following the typical rules of stride-one access, recurrence free, and limited conditional logic. To accomplish good maintainability and portability we do not explicitly program this level of parallelization and leave it to a compiler to decide if and when to use vectorization. This usually requires scalar operations be isolated to separate loops so the remaining work is available for vectorization.  

\subsubsection{OpenMP Threading} \label{opemp_design}

The benefits of hybrid application scaling to high thread counts was discussed in \S \ref{introduction}.  We avoid the bottle-necks of parallel loop-based OpenMP by arranging WOMBAT such that only one OpenMP parallel region is present for the duration of execution.  This design presents the threaded region as a set of completely independent processes, which mimics the parallelism of MPI. We refer to this approach as SPMD OpenMP (or single program-multiple data OpenMP) \citep[see also][]{cug_knl_mpich}.

\begin{figure}
\centering
\includegraphics[height=.5\textheight]{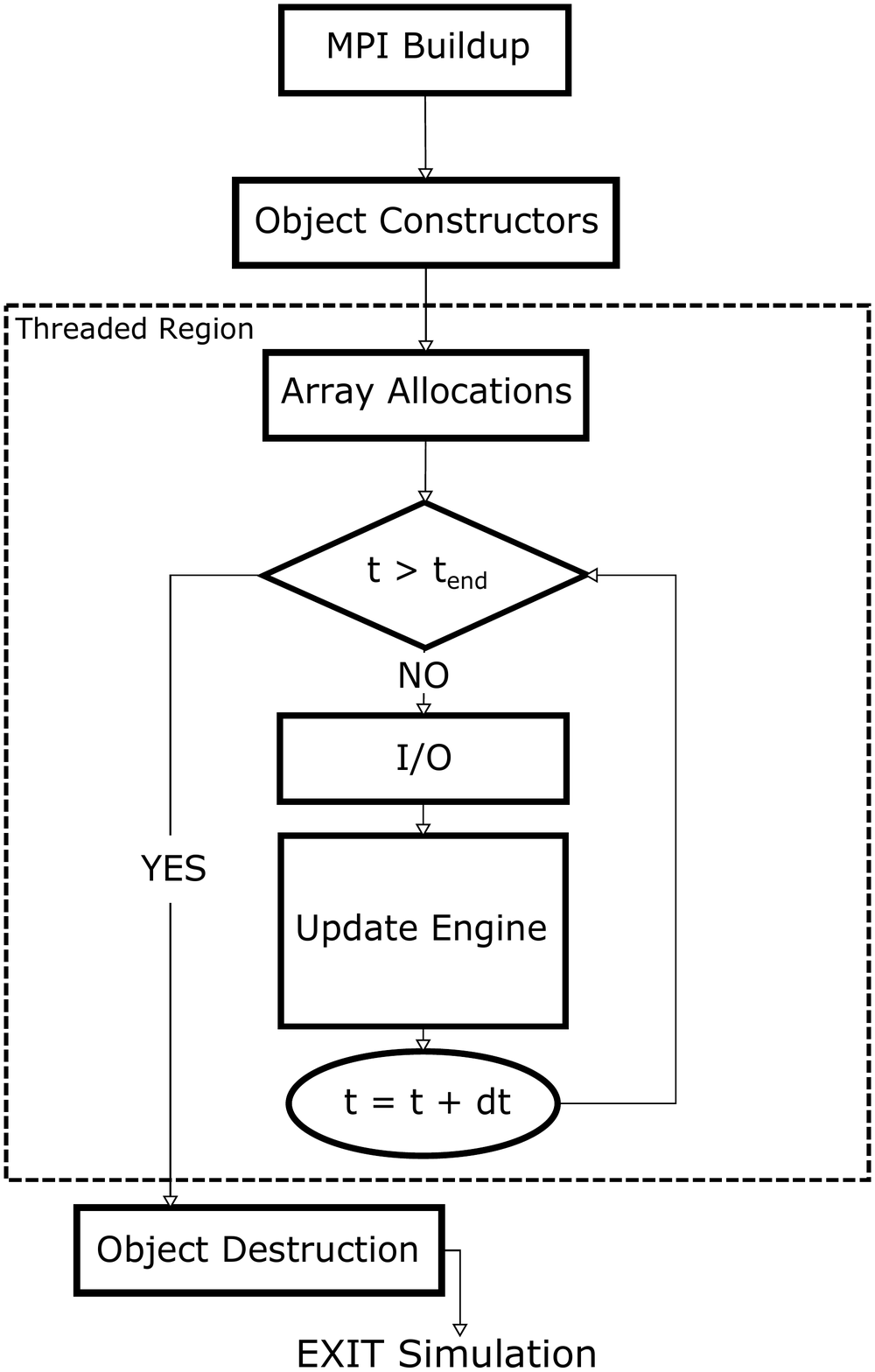}
\caption{Flow diagram for the main driver in WOMBAT.  The scope of the \emph{single} OpenMP parallel region is outlined with a dashed box.} \label{fig:driver}
\end{figure}

To illustrate this design, we show a flow diagram of the main driver in WOMBAT in Figure \ref{fig:driver}.  A section including MPI initialization and base object construction is the only work by the main thread outside the parallel region.  After that, every portion of WOMBAT is executed by all threads collaboratively.  This includes array allocation/paging, computation, communication, and even I/O.

\subsubsection{MPI} \label{mpi}

To allow fine-grained work-communication overlap, the Patch design results in a larger number of small messages, compared to traditional approaches which typically use fewer monolithic boundary exchange messages. This shifts the communication sensitivity of WOMBAT from simply bandwidth to bandwidth and message rate, depending on the number and size of Patches. A Domain decomposed into Patches will generate more MPI messages and a higher aggregate amount of data moved, especially in 3d, due to added corner and edge boundaries.  This extra communication will later be leveraged for communicating load information and changes to Patch ownership.  Furthermore, in the SPMD OpenMP approach every thread can participate in the MPI communication using \textbf{\mbox{MPI\_THREAD\_MULTIPLE}}.\par
Most of the MPI communication in WOMBAT uses MPI-RMA (\emph{e.g.}, \textbf{\mbox{MPI\_Put()}}, \textbf{\mbox{MPI\_Get()}}), because of the low overhead possible with a proper MPI-RMA implementation. MPI-RMA was added to the MPI standard primarily to give users direct access to the Remote Direct Memory Access (RDMA) features available on most HPC networks (interconnects).  PUT and GET operations in MPI-RMA are inherently non-blocking, and excellent overlap of computation and communication is possible on networks that also support network-offloading.  While the semantics of MPI-RMA allow for these performance characteristics, many MPI libraries today implement MPI-RMA using two-sided communication \citep[\emph{e.g.,}][]{c86f5a6e77c8463b8f41ed7e67bb822c}.  This adds overhead and reduces the chances for overlap, leaving MPI-RMA practically unusable for an HPC application.  Recent work in MVAPICH \citep{mvapich1,mvapich2,mvapich3} and OpenMPI \citep{Hjelm:2014:OOO:2642769.2642792,Hjelm:2016:EOP:2966884.2966890} has corrected that issue on both Infiniband and Cray networks.  In Cray MPICH, MPI-RMA is now based on the low level DMAPP library specifically designed for optimal one-sided communication on Gemini and Aries interconnects.  This implementation has very low overhead, tuned to utilize the network-offload (Block Transfer Engine or BTE) capability on Cray XE/XC systems.\par
Message rate requirements and the SPMD design make it critically important that multi-threaded MPI-RMA in a given MPI library performs well. During the design of WOMBAT we found that no MPI implementation really achieved the performance that should be possible. This is because most MPI implementations (including Cray MPICH at the time) use a global lock to provide thread safety \citep[see also][]{dosanjh2016rma}. This serializes all MPI calls, and more importantly, most work in the user code around those MPI calls. Through a co-design approach we have optimized Cray MPICH for high performance and thread scalable  MPI-RMA communication (see \S \ref{mpich_rma}).  We refer to this new capability as ``thread-hot RMA''. An initial version is available to Cray users starting with Cray MPICH 7.3.4 with additional enhancements from the work presented here available in an upcoming releases.  Other MPI libraries are also pursuing optimizations for \textbf{\mbox{MPI\_THREAD\_MULTIPLE}} that will make our design performance portable beyond Cray systems \citep{amer2015mpi,vaidyanathan2015improving}.

\subsubsection{Asynchronous I/O} \label{io}

WOMBAT uses a custom asynchronous I/O (AIO) library that allows for overlap of simulation progression and data writing. If all I/O data can be buffered, data can be written out with almost no impact on execution time.  Some portion of I/O work is done blocking if buffers are made smaller, which reduces overall performance.

AIO is implemented as a set of specialized ranks dedicated to receiving (or sending for read operations) data from a client set of worker ranks.  All threads in the worker ranks package data into I/O buffers.  Non-blocking communication is used to move data to AIO server ranks, and the AIO ranks then write data out as it comes in. The full system can be tuned for I/O and overlap performance by adjusting the total number of AIO server ranks.


\section{Performance} \label{performance}

\begin{deluxetable*}{ccccccc}
\tablecolumns{7}
\tabletypesize{\scriptsize}
\tablecaption{Test platforms used in the performance studies and their characteristics. Note: We mostly use 64 cores on KNL systems to make scaling studies simpler multiples from lower core counts. \label{tab:test_systems}}
\tablehead{
    \colhead{System Title}
  & \colhead{Architecture}
  & \colhead{Interconnect}
  & \colhead{Topology}
  & \colhead{CPU}
  & \colhead{VL [bits]}
  & \colhead{Cores per Node}
}
\startdata
Blue Waters & Cray XE & Cray Gemini & 3d torus & AMD Opteron$^{TM}$ 6276 ``Interlagos'' @ 2.3 GHz & 256 & 16 \\
XE$_{IL}$  & Cray XE & Cray Gemini & 3d torus & AMD Opteron$^{TM}$ 6281 ``Interlagos'' @ 2.5 GHz & 256 & 16 \\
XC$_{BDW}$  & Cray XC & Cray Aries & dragonfly & Intel\textsuperscript{\textregistered} Xeon\textsuperscript{\textregistered} E5-2695 ``Broadwell'' @ 2.5 GHz & 256 & 36 \\
XC$_{KNL}$  & Cray XC & Cray Aries & dragonfly & Intel\textsuperscript{\textregistered} Xeon Phi$^{TM}$ 7250 ``KNL'' @ 1.4 GHz & 512 & 68 \\
\enddata
\end{deluxetable*}

We measure the performance of WOMBAT for 3d MHD calculations using the directionally un-split MHDTVD solver described in \S \ref{methods}.  Single node tests focus on the impact of vectorization on overall execution and the parallel efficiency of the SPMD OpenMP technique.  Multi-node tests at scale measure the performance of the full suite of parallelization strategies, including off-node communication, and how they interact. We stress that the problems sizes used for most experiments presented here were selected to show overheads in WOMBAT, and in particular communication.  This also closely follows real-world simulations run on production systems.\par
Table \ref{tab:test_systems} summarizes the platforms used for performance experiments.  We used \emph{Blue Waters} (Cray XE) at the NCSA for very large weak scaling studies.  The remaining systems are internal configurations at Cray Inc.  We use the Cray Compiler (CCE) in all experiments.  Table \ref{tab:test_notes} shows specific test information about each processor.  We used the new Cray MPICH library with the ``thread-hot RMA'' feature in all tests unless otherwise noted. \par
All experiments involving KNL were run with nodes configured in so-called ``quadrant'' Non-Uniform Memory Access (NUMA) mode with high bandwidth memory (on package) configured as a 16 GB L3 cache.

\begin{deluxetable*}{ccc}
\tablecolumns{3}
\tabletypesize{\scriptsize}
\tablecaption{Processor compilation and placement notes. \label{tab:test_notes}}
\tablehead{
    \colhead{CPU}
  & \colhead{Compilation Flags}
  & \colhead{Notes}
}
\startdata
Interlagos & -O vector3 -h preferred\_vector\_width=256 & Only one thread/process per floating point unit \\
Broadwell & \emph{defaults} & Only one thread/process per core (no hardware threads used) \\
KNL & \emph{defaults} & Only one thread/process per core (no hardware threads used) \\
\enddata
\end{deluxetable*}

\subsection{Single Node Performance by Architecture}

\subsubsection{SIMD Scaling}

\begin{figure}
	\centering
	\includegraphics[height=.25\textheight]{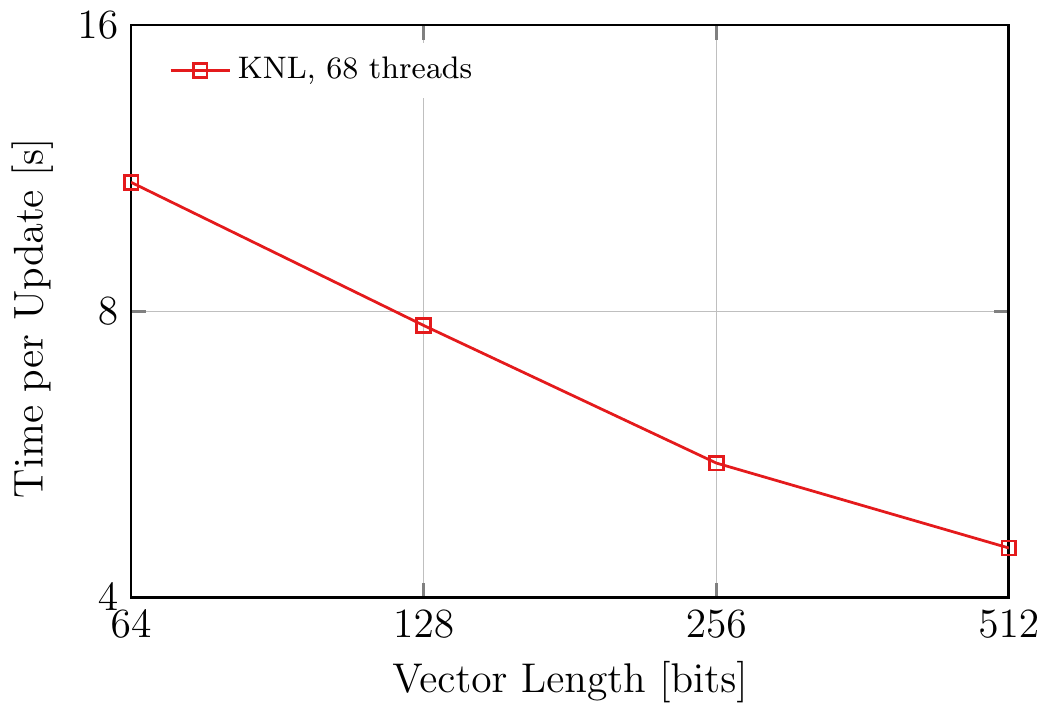}
	\caption{Time per update in seconds for increasing VL (strong scaling) on KNL with 68 threads.  }
	\label{fig:vl_scaling}
\end{figure}

We measure the impact of increasing vector length (VL) from 64 to 512 bits on KNL\footnote{To vary the type of vectors, we use the CCE compiler flag \textbf{``-h preferred\_vector\_width=X''}, where \textbf{X} = \{128, 256, 512\}. For scalar 64 bit vectors we use \textbf{``-O vector0''}.}.  The problem size is a 17x4$^2$ Domain of Patches each with 48$^3$ zones updated by a single MPI rank with 68 threads.  \par
Figure \ref{fig:vl_scaling} shows the strong scaling with increasing VL on a single node of the XC$_{KNL}$ system.  The time to perform a single time step update is reduced approximately by a factor of 2 going from 64 and 256 bit vectors.  The final step to 512 bit vectors continues to show improved performance, but the effect has been reduced to only a $\simeq$ 19\% speedup.  Factors that affect the speedup from vectorization are the amount of vector versus scalar code executed, the efficiency of the vector code generated by the compiler, and the memory bandwidth available to provide data to the cores.  Overall, vectorization speeds up WOMBAT by almost a factor of 2.5X on KNL processors.  The speedup is roughly consistent with Amdahl's Law, assuming the fraction of execution  time benefitting from parallelization $p \approx 0.65$.

Broadwell has accessible hardware performance counters for floating point operations that can be measured with a number of performance tools, such as PAPI or CrayPAT.  Table \ref{tab:vector_instr} shows the quality of vectorization relative to a scalar build of WOMBAT.  We also show the breakdown of scalar and vector operations for the Intel and GNU compilers.  Vectorization with CCE reduces total double precision (DP) floating point instruction count by $\sim$ 71\%.  79\% of all floating point operations are vector.  Intel produces a similar amount of vector instructions but also lower performance.  The performance difference is due to much lower translation lookaside buffer (TLB) utilization despite using 2 Megabyte huge pages.  We intend to file a performance bug with Intel on this issue and it will be corrected in later releases.  The GNU compiler is unable to produce any vector instructions.

\begin{deluxetable*}{ccccc}
\tablecolumns{5}
\tablecaption{Effect of vectorization on floating point instruction count on Broadwell for a single thread running a 2$^3$ x 32$^3$ problem for 100 time steps. \label{tab:vector_instr}}
\tablehead{
    \colhead{Compiler}
  & \colhead{10$^9$ Scalar Ops.}
  & \colhead{10$^9$ 256 B SIMD Ops.}
  & \colhead{sec / update}
  & \colhead{\% of DP Peak}
}
\startdata
CCE 8.5.4 (scalar) & 327 & 0 & 1.87 & 6.6 \\
CCE 8.5.4 & 20 & 74 & 0.96 & 14.1 \\
Intel 17.0.1.132 & 15 & 75 & 1.67 & 8.1 \\
GNU 6.2.0 & 320 & 0 & 1.84 & 6.6 \\
\enddata
\end{deluxetable*}

\subsubsection{Thread Scaling} \label{thread_scaling}

\begin{figure}
	\centering
	\includegraphics[width=0.48\textwidth]{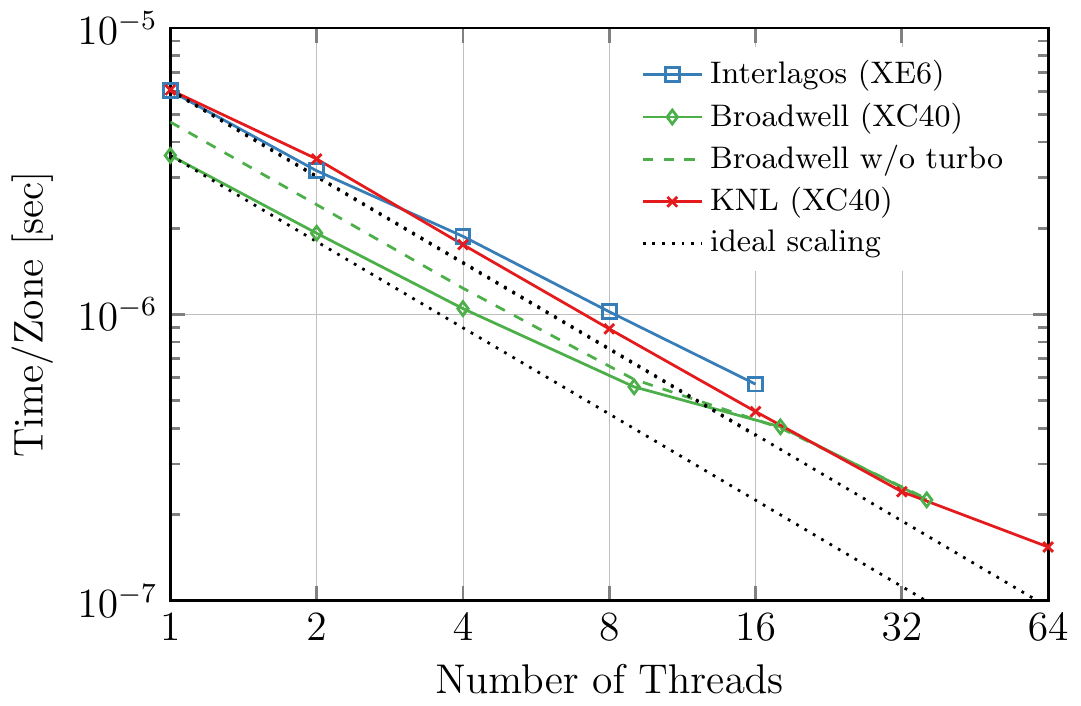}
	\caption{Execution time per zone over number of threads (strong scaling) for Interlagos (blue), Broadwell (green) and KNL (red).}
	\label{fig:thread_strong_scaling}
\end{figure}

\begin{deluxetable}{ccc}
\tablecolumns{3}
\tablecaption{Number of Patches in each direction of the Domain and number of zones per Patch used by each architecture in the thread strong scaling test. \label{tab:thread_scaling}}
\tablehead{
    \colhead{CPU}
  & \colhead{Domain Size}
  & \colhead{Patch Size}
}
\startdata
Interlagos & 8x7x4 & 40$^3$ \\
Broadwell & 12$^2$x8 & 32$^3$ \\
KNL & 8$^3$ & 48$^3$ \\
\enddata
\end{deluxetable}

We show the thread strong scaling speedup of WOMBAT on the three architectures presented here in Figure \ref{fig:thread_strong_scaling} with the problem sizes given in Table \ref{tab:thread_scaling}. The code shows excellent speedup with threads on all the architectures.  The speedup from threads on KNL is $\sim$ 40X at 68 threads. The ``turbo'' on Broadwell increases performance for small thread numbers (green line versus dashed green line). On both Interlagos and Broadwell there is notable loss of scaling once the process has threads spanning beyond a single NUMA domain. On Interlagos this is at 4 threads, because each Interlagos processor is made up of two ``Bulldozer'' modules for a total of four NUMA nodes on a dual socket XE node.  A KNL node configured in quadrant mode has only a single NUMA node, and the deviation from ideal scaling is moved to much higher thread counts.  Profiles showed that the cost of thread synchronization rising at these thread counts, but there is the additional factor of a finite amount of bandwidth available on the processor. Both of these account for most of the reduction in performance from ideal.

\subsection{Performance at Scale by Architecture}

Off-node components of an application, such as network latency and bandwidth, can modify its behavior and how it should be tuned.  We present a multi-node Patch size optimization study for WOMBAT, and we also demonstrate the weak and strong scaling capabilities of WOMBAT out to large node counts.

\subsubsection{Patch Size Optimization} \label{patch_optimization}

\begin{figure*}
	\centering
	\includegraphics[width=0.45\textwidth]{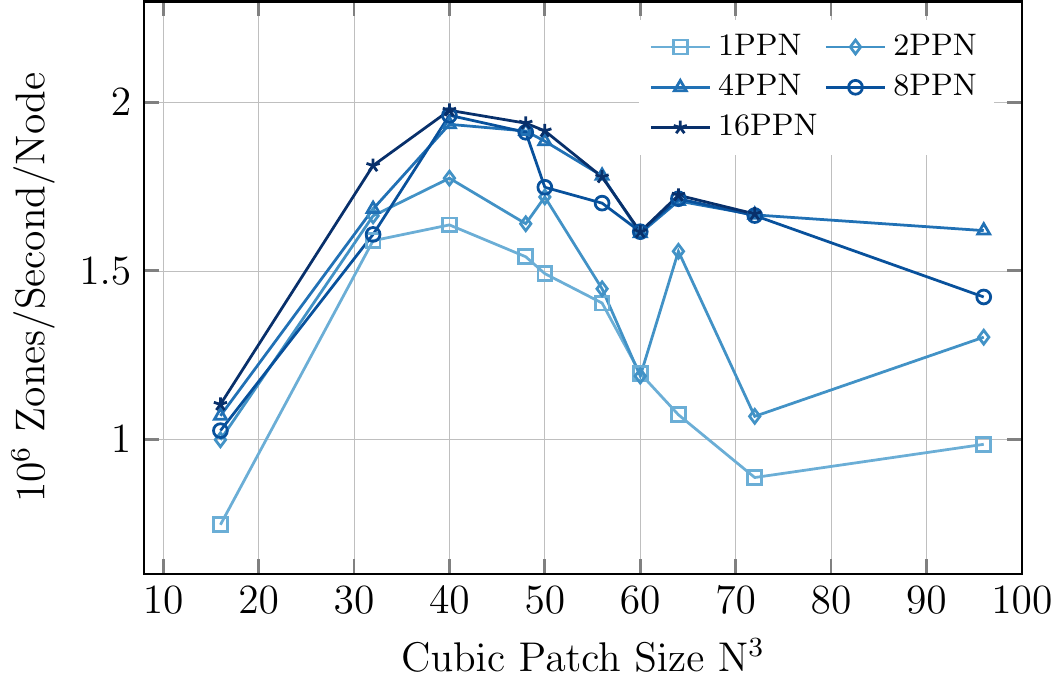}
	\includegraphics[width=0.45\textwidth]{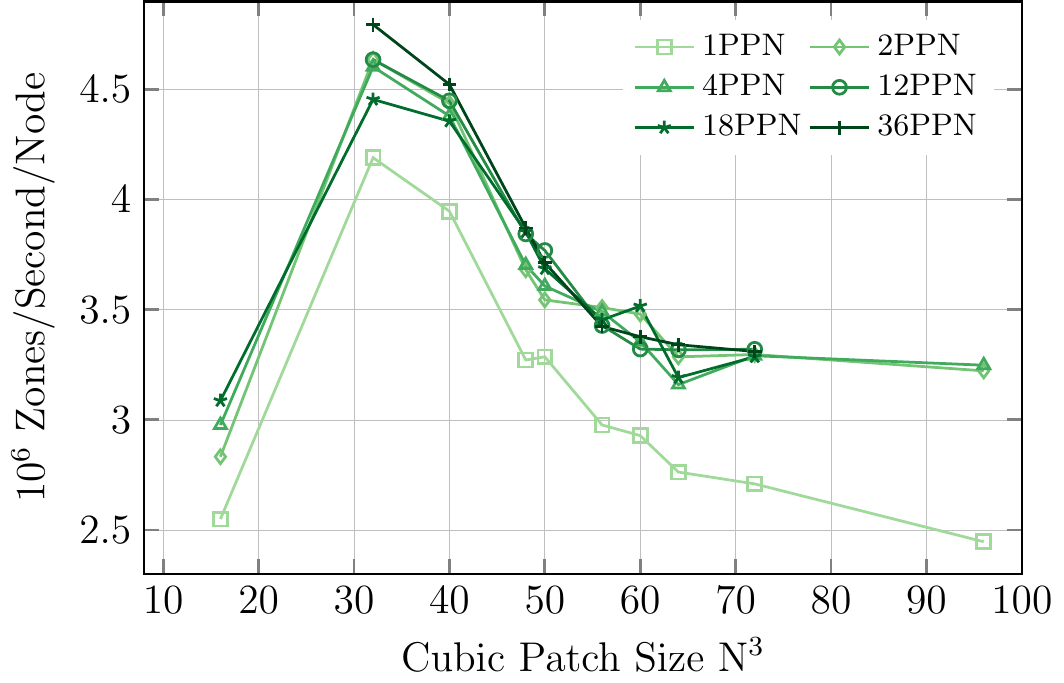} \\
	\includegraphics[width=0.45\textwidth]{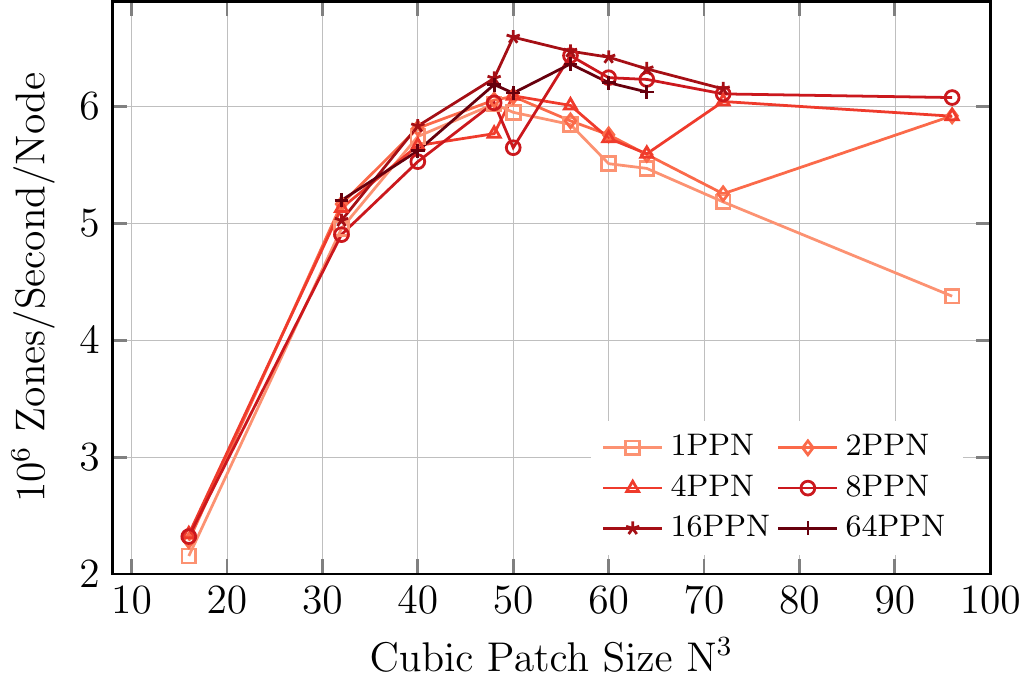}
	\caption{Performance in Million zones per second per node over number of zones
along one patch dimension. We show Interlagos (blue, \emph{top left}) from the XE$_{IL}$ system, Broadwell (green,
\emph{top right}) from the XC$_{BDW}$ system and KNL (red, \emph{bottom}) from the XC$_{KNL}$ system at different values of MPI processes per node (PPN). } \label{fig:patchopt}
\end{figure*}

In \S \ref{decomposition} we described how Patches are logically assembled to produce any grid size per rank.  We study performance with Patch size also allowing the mixture of MPI ranks to OpenMP threads to vary at a scale of 27 nodes.  27 nodes is used because any configuration of MPI ranks to threads at that scale will have unique neighbors in 3d.  This ensures that the MPI work is saturated and performance is not skewed.  We use ``PPN'' to denote the number of MPI processes per node with threads placed on all cores. The total number of zones across each Patch size was held approximately constant within a system type. We chose the problem setup  so update-times were held at $\simeq$ 10 seconds. This is sufficiently large to expect throughput values to be near their absolute peak but still include overhead sensitivity. \par
Figure \ref{fig:patchopt} shows the throughput on each architecture given by the number of zones per second each node can update. We can identify the maximum throughput each system can achieve for these problem setups.  XE$_{KNL}$ nodes are able to update $\gtrsim$ 6x10$^6$ zones/sec/node at peak compared $\gtrsim$ 4.5x10$^6$ zones/sec/node on XC$_{BDW}$ and 2x10$^6$ zones/sec/node on XE$_{IL}$.  This demonstrates the ability of our approach to adapt to the unique many-core design of KNL. \par
The optimal Patch size is not uniform across systems.  XC$_{BDW}$ has the smallest optimal Patch size of 32$^3$.  The optimal Patch size for XE$_{IL}$ is 40$^3$, and XC$_{KNL}$ has an optimal Patch size $\simeq$ 50$^3$.  The performance on either side of the optimal size drops off but not by the same amount on each system.  Patch sizes smaller than the optimal size have lower performance due to reduced vectorization efficiency, and larger sizes have lower performance due to spilling out of L3 on both Interlagos and Broadwell.  KNL has the largest SIMD vector size, which explains why it has the largest optimal Patch size.  Overheads appear to affect Interlagos more than Broadwell.  This favors slightly larger Patches on Interlagos despite having the same SIMD vector length as Broadwell. The cache blocking properties of the Patch design discussed in \S \ref{cache_blocking} no longer function as intended for large Patches.  On KNL the performance loss is not as dramatic, with exceptions at high thread count.  This is due to the 16 GB L3 cache.\par
Both XE$_{IL}$ and XC$_{BDW}$ show lowest performance with only a single rank per node packed with threads.  This is largely due to the NUMA issues discussed in \S \ref{thread_scaling}.  In both of these cases, once the number of MPI processes per node matches the number of NUMA nodes on a node there is very little spread in performance at the optimal Patch size.  In the case of XC$_{KNL}$ performance does not vary much until the Patch size is 50$^3$ or greater.  While the absolute best performance on XC$_{KNL}$ is with 16 PPN ($\sim$ 6.5x10$^6$ zones/sec/node), there is still significant performance at 4 PPN ($\sim$ 6x10$^6$ zones/sec/node).  At a high level, a fixed grid calculation should not perform any different exchanging ranks for threads if both MPI and OpenMP are well implemented and hardware limitations are not present.  The SPMD design in WOMBAT nearly achieves this.\par

\subsubsection{Weak Scaling}

\begin{figure*}[h!]
	\centering
	\includegraphics[width=0.45\textwidth]{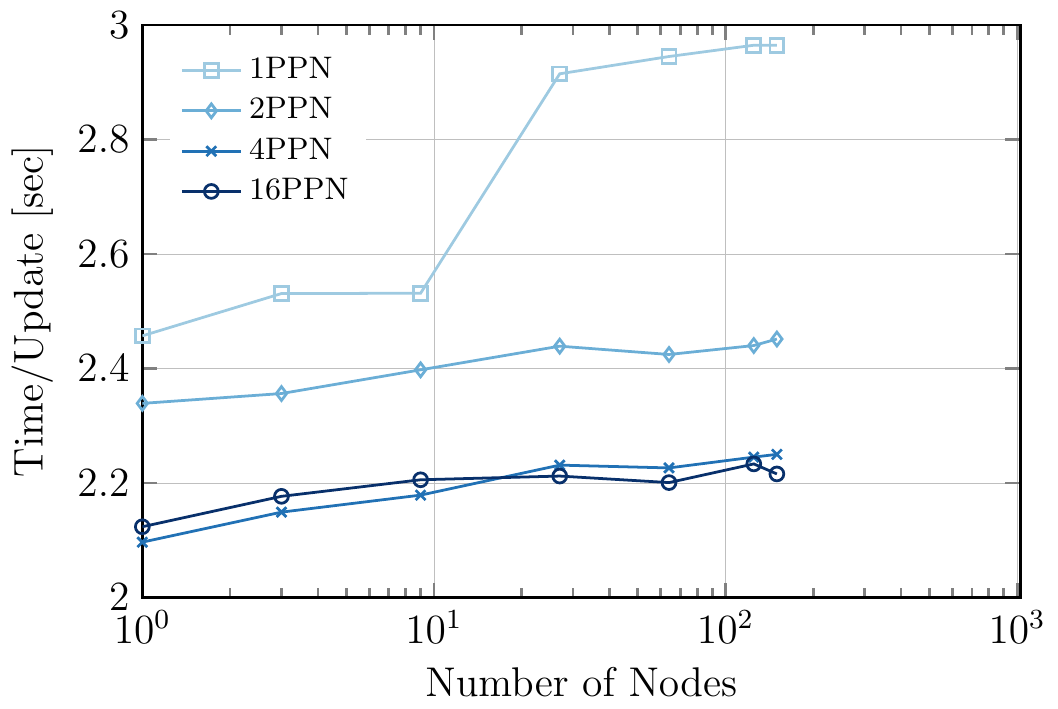}
	\includegraphics[width=0.45\textwidth]{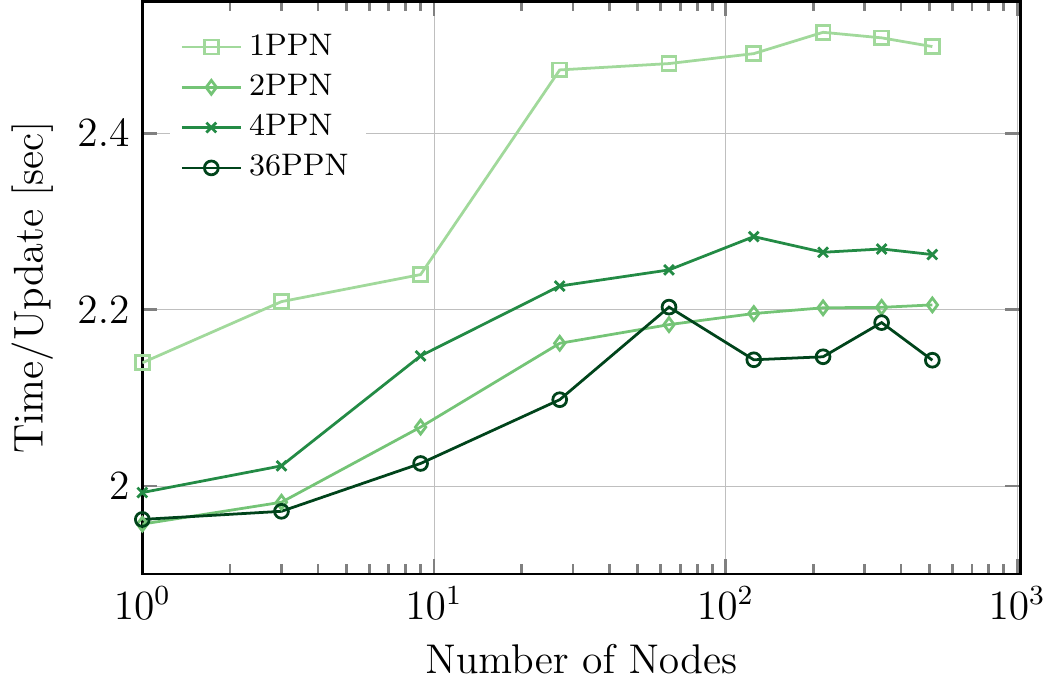}\\
	\includegraphics[width=0.45\textwidth]{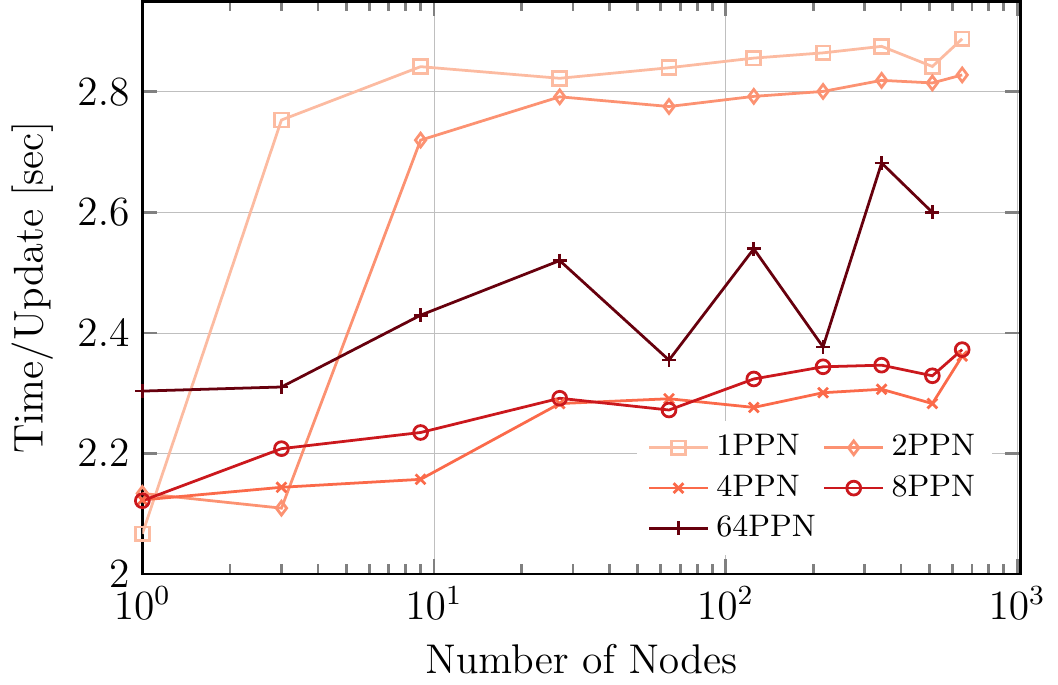}
	\caption{Time per update over number of full nodes (weak scaling) for XE$_{IL}$ (blue, \emph{top
	left}), XC$_{BDW}$ (green, \emph{top right}) and XC$_{KNL}$ (red, \emph{bottom}) at different values of MPI processes per node (PPN).}
	\label{fig:weak_scaling}
\end{figure*}

\begin{deluxetable}{ccc}
\tablecolumns{3}
\tablecaption{Weak Scaling Test Setup: The number of Patches in each direction of the Domain and number of zones per Patch used for each system. \label{tab:weak_scaling}}
\tablehead{
    \colhead{System}
  & \colhead{Patches per Node}
  & \colhead{Patch Size}
}
\startdata
XE$_{IL}$ & 4x4x4 & 40$^3$ \\
XC$_{BDW}$ & 6x6$^2$ & 32$^3$ \\
XC$_{KNL}$ & 8x4$^2$ & 48$^3$ \\
\enddata
\end{deluxetable}

\begin{figure*}
	\centering
	\includegraphics[height=.4\textheight]{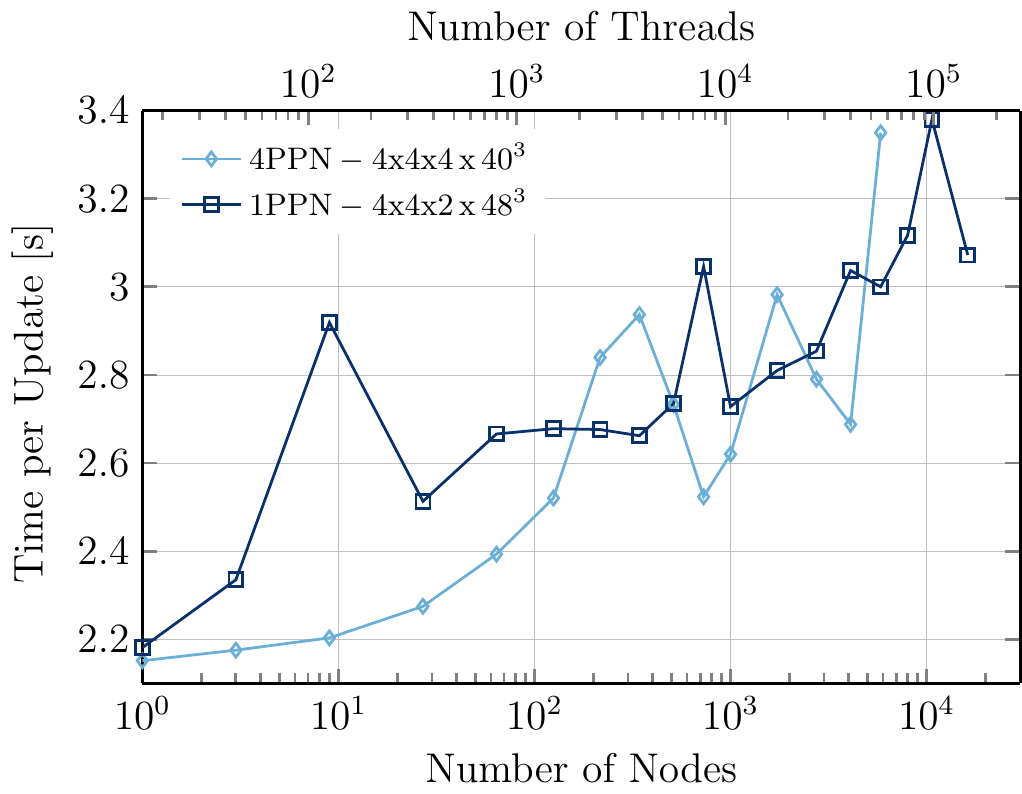}
	\caption{Weak scaling on Blue Waters for two different values of MPI processes per node (PPN) and problem size.}
	\label{fig:bw_weak_scaling}
\end{figure*}

Figure \ref{fig:weak_scaling} shows the weak scaling on three architectures at different values of PPN for the problem sizes given in Table \ref{tab:weak_scaling}. For XC$_{IL}$ and XE$_{BDW}$ the best performance and scaling is closely matched between a single MPI rank per NUMA node or pure MPI, which follows the conclusions in \S \ref{patch_optimization}.  The XC$_{KNL}$ systems has best performance and scaling at 4 and 8 PPN.  Relative to a \emph{single node}, XE$_{IL}$ has a 93\% efficiency up to 150 nodes at 4 PPN, and XC$_{BDW}$ at 2 PPN has a 87\% efficiency up to 512 nodes.  Remarkably XC$_{KNL}$ has 89\% efficency at 648 nodes (41,472 threads) with 4 PPN relative to a single node run. \par
The lowest performance at scale on all systems is with a single rank per node.  The increases in update times are due to increasing amounts of off-node communication a rank encounters and imperfect overlap of communication with computation.  Between 3 and 27 nodes, off-node communication cost is saturated, and update times are nearly flat for larger node counts. We conclude that WOMBAT has excellent weak scaling on all systems with the optimal configuration despite the relatively small problem size chosen. \par
We show weak scaling on Blue Waters in Figure \ref{fig:bw_weak_scaling} for two values of PPN and problem sizes.  We limit each run to just 10 time steps. The 1 PPN runs scale out to 16,224 nodes (259,584 threads) with a world grid containing $\simeq$ 66 billion zones.  WOMBAT scales well with 60\% efficiency at 16,224 nodes for 1 PPN and 75\% efficiency at 4,096 nodes for 4 PPN relative to single node runs.

There are several spikes of increased update times.  We hypothesize it is due to network contention with both other running applications and with WOMBAT itself on the very large 3d torus topology.  Runs on the smaller dedicate XE$_{IL}$ system do not show these features.  Going forward, we intend to make use of the topology-aware scheduling capability provided by the NCSA, and we expect this to improve performance and reduce the contention opportunities.

The right panel of Figure \ref{fig:strong_scaling} shows the impact of the ``thread-hot RMA'' capability that will be included in an upcoming release of Cray MPICH. We show weak scaling on XC$_{KNL}$ to 125 nodes with 4 PPN using the same problem setup for weak scaling described above.  This feature produces a 17\% speedup over Cray MPICH 7.3.1.

\subsubsection{Strong Scaling}

\begin{figure*}
	\centering
	\includegraphics[width=.46\textwidth]{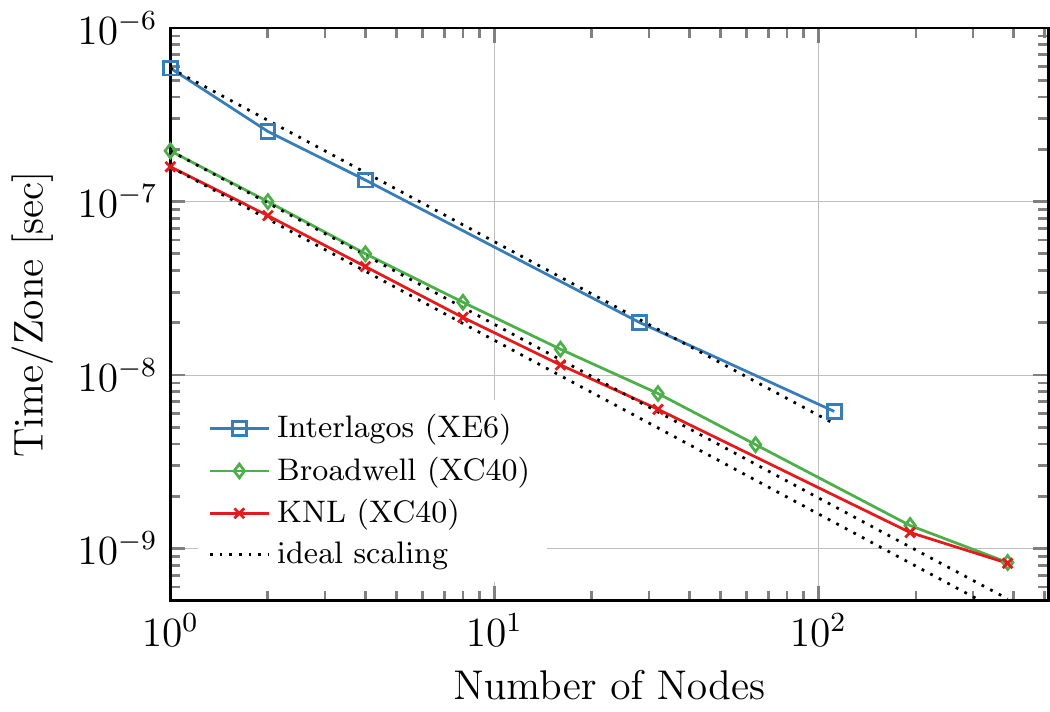}
	\includegraphics[width=.45\textwidth]{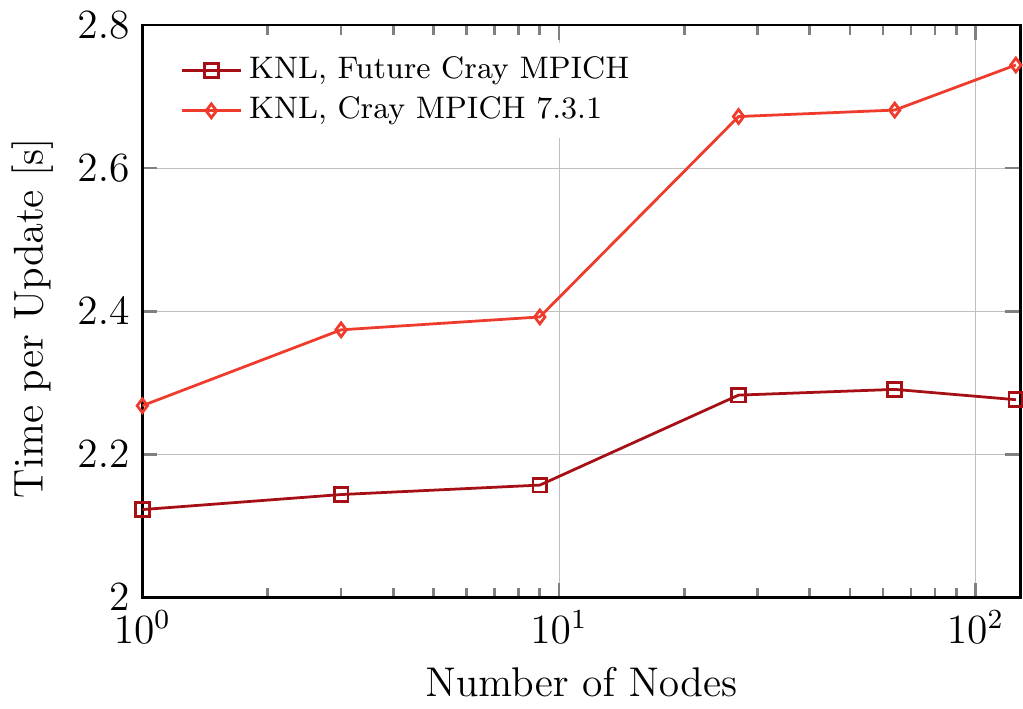}
	\caption{\emph{Left}: Update time per zone over number of nodes (strong scaling) for XE$_{IL}$ (blue), XC$_{BDW}$ (green) and XC$_{KNL}$ (red) at constant problem size. \emph{Right}: Update time in seconds over number of nodes (strong scaling) for Cray MPICH 7.3.1 (light red) and new lock optimized Cray MPICH (dark red) on XC$_{KNL}$ with 68 threads.}
	\label{fig:strong_scaling}
\end{figure*}

\begin{deluxetable}{cccc}
\tablecolumns{4}
\tabletypesize{\scriptsize}
\tablecaption{Strong scaling test setup. \label{tab:strong_scaling}}
\tablehead{
    \colhead{System}
  & \colhead{PPN}
  & \colhead{World Grid Patches}
  & \colhead{Base Patch Size}
}
\startdata
XE$_{IL}$ & 4 & 14x8x12 & 40$^3$ \\
XC$_{BDW}$ & 2 & 48x12$^2$ & 32$^3$ \\
XC$_{KNL}$ & 2 & 32x16x12 & 40$^3$ \\
\enddata
\end{deluxetable}

The left panel of Figure \ref{fig:strong_scaling} shows the strong scaling of WOMBAT.  We defined the problem size for each system so that the time per update is limited to $\sim$ 60 seconds (see Table \ref{tab:strong_scaling}).  Performance closely follows the theoretical speedup over 2 orders of magnitude in node count, with XC$_{BDW}$ showing a 236X speedup at 384 nodes. The deviations are due to overheads exposed as update times at scale fall at or below 0.3 seconds.  We reduced the Patch size on XC$_{BDW}$ by half and on XC$_{KNL}$ by a quarter at the largest scale, for example.

\subsubsection{Thread Scheduling at Scale}

\begin{figure}
	\centering
	\includegraphics[width=0.45\textwidth]{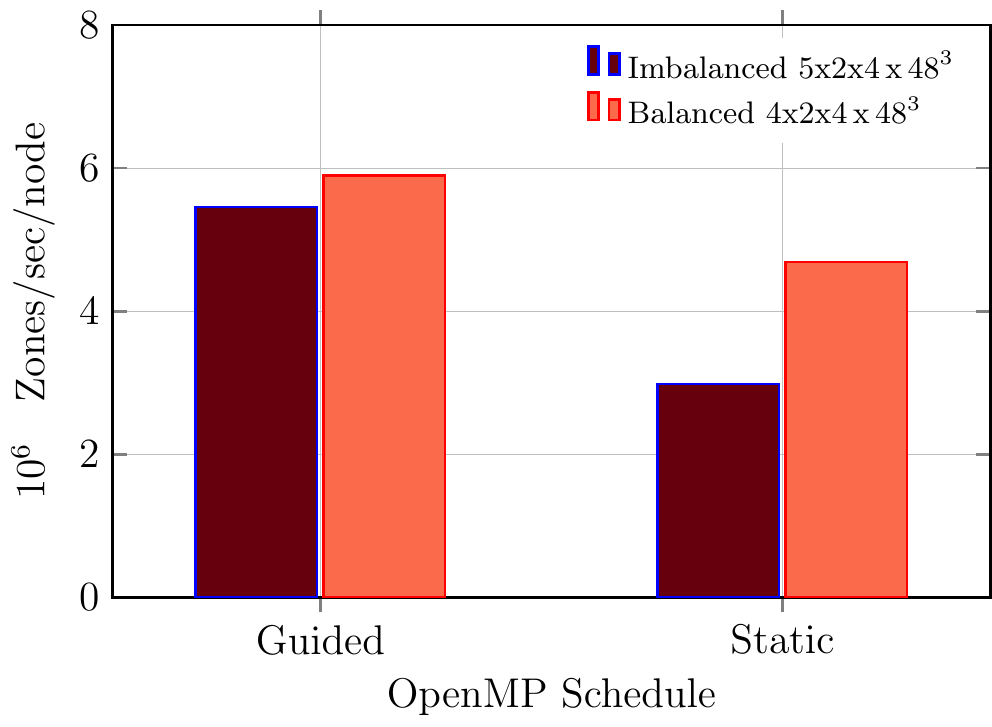}
	\caption{Comparison between the OpenMP \textbf{GUIDED} loop schedule and the \textbf{STATIC} schedule as applied to the DomainSolver class Patch update loop for a balanced (even work per thread) and imbalanced problems.} \label{fig:thread_imbalance}
\end{figure}

We show the impact of OpenMP scheduling in Figure \ref{fig:thread_imbalance}.  In this experiment we modified the thread scheduling for just a single loop in the Update Engine  (see \S \ref{update_engine}) that drives updates over Patches.  We run a balanced (Patch count divides evenly into thread count) and an imbalanced problem with the modified code and original code. 4 MPI ranks per node on 27 nodes each with 16 threads update 40 (imbalanced) or 32 (balanced) Patches, each with 48$^3$ zones on the XC$_{KNL}$ system.  Both problems show near optimal throughput with the \textbf{GUIDED} schedule and lower performance with \textbf{STATIC}. Our SPMD OpenMP design has very few thread barriers, and using a \textbf{STATIC} schedule assumes threads are roughly synchronized to be efficient.


\section{Design Details} \label{design_details}

\subsection{Update Engine} \label{update_engine}

\begin{figure}
	\centering
\includegraphics[height=.5\textheight]{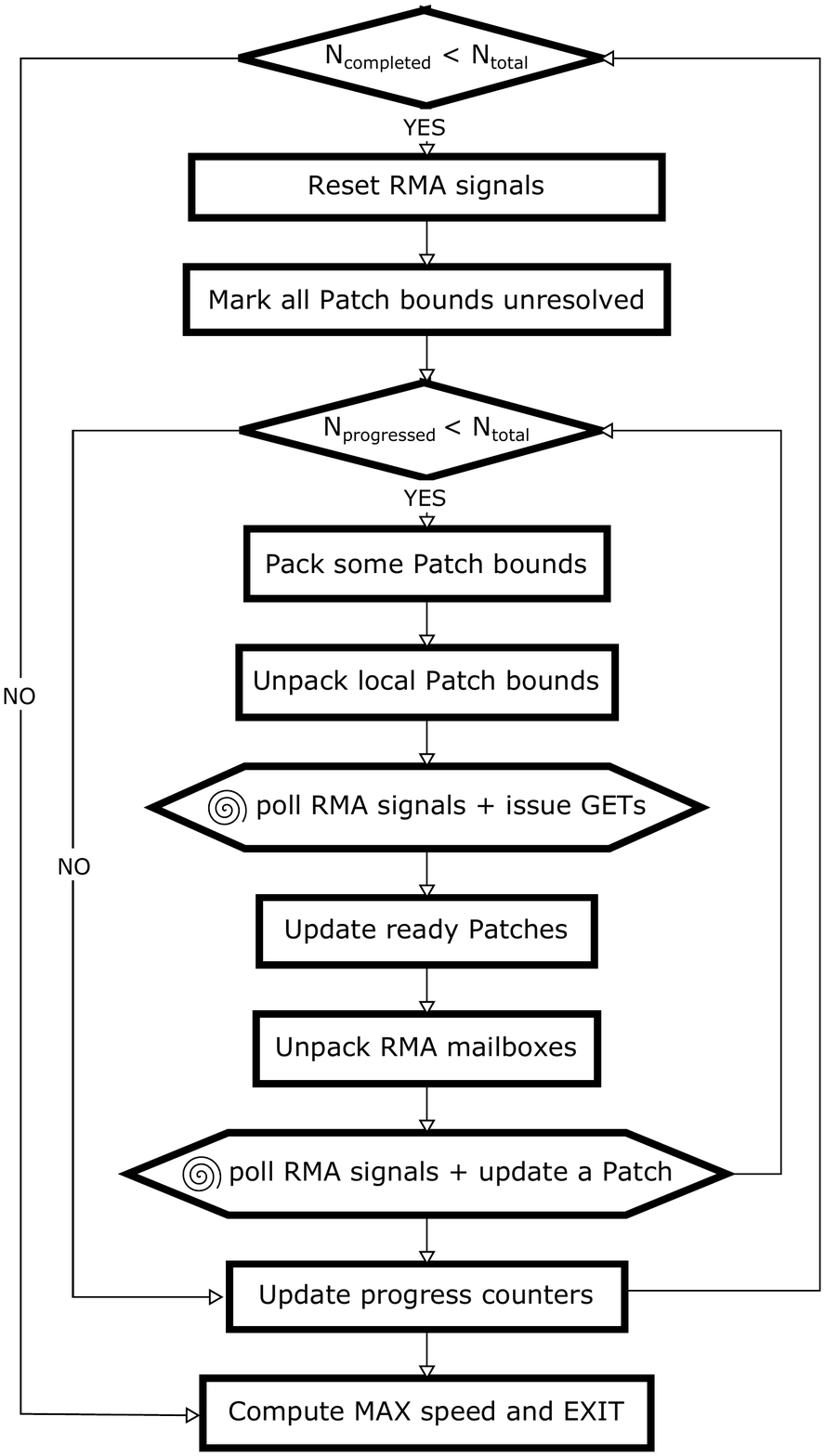}
\caption{Generalize communication/update engine.} \label{fig:update_engine}
\end{figure}

In our design, the Update Engine is responsible for scheduling computation and communication across threads for any solver.  It accepts a Domain and iterates over Patches, exchanging messages and partitioning the update work through the specified solver until all Patches report back as completed. To allow for iterative or sub-cycling solvers, it is not necessary that a Patch be updated completely for the current time step after only a single pass through a solver. \par
Figure \ref{fig:update_engine} shows a schematic of the Update Engine.  It is contained inside the higher level OpenMP parallel region shown in Figure \ref{fig:driver}, and all threads call it with the same input data and requested solver class.  The outer \emph{while} loop allows for iterative solvers that require any arbitrary number of passes (including messaging) to complete for a time step.  The inner \emph{while} loop contains the work necessary to drive both communication and updates through the requested solver.  The work includes packing (unpacking) boundary data into (out of) contiguous buffers, used for either local copies within a rank or MPI transfers between ranks.  There is additional work for signaling and data transfer between MPI ranks and updating Patches with resolved boundaries through the requested solver.  The DomainSolver class manages all book-keeping related to marking individual boundaries for any affected Patch as resolved/unresolved.  It also tracks grids within a Patch as incrementally or completely updated.\par
An important optimization in this design relates to how boundary data is exchanged between Patches contained in the same MPI rank.  An instance of the Patch class includes a buffer for \emph{incoming} boundary data (there is no matching buffer for outgoing data).  This buffer is only used for boundary data coming from another local Patch.  With the Domain class, data destined for a local Patch is directly packed into the buffer of the destination Patch.  The buffer is later unpacked, along with any non-local boundary data, into the Patch grid boundary zones.  This optimization takes advantage of the shared memory aspect of OpenMP, completing local Patch boundary exchange without calling MPI or excessive buffering.  Some buffering is necessary to minimize contention between threads attempting to progress the same Patch.  A single node run can completely avoid calling MPI with this feature by using threads on all cores.

\subsubsection{MPI-RMA Engine} \label{rma_engine}

\begin{figure*}
\centering
\includegraphics[height=.25\textheight]{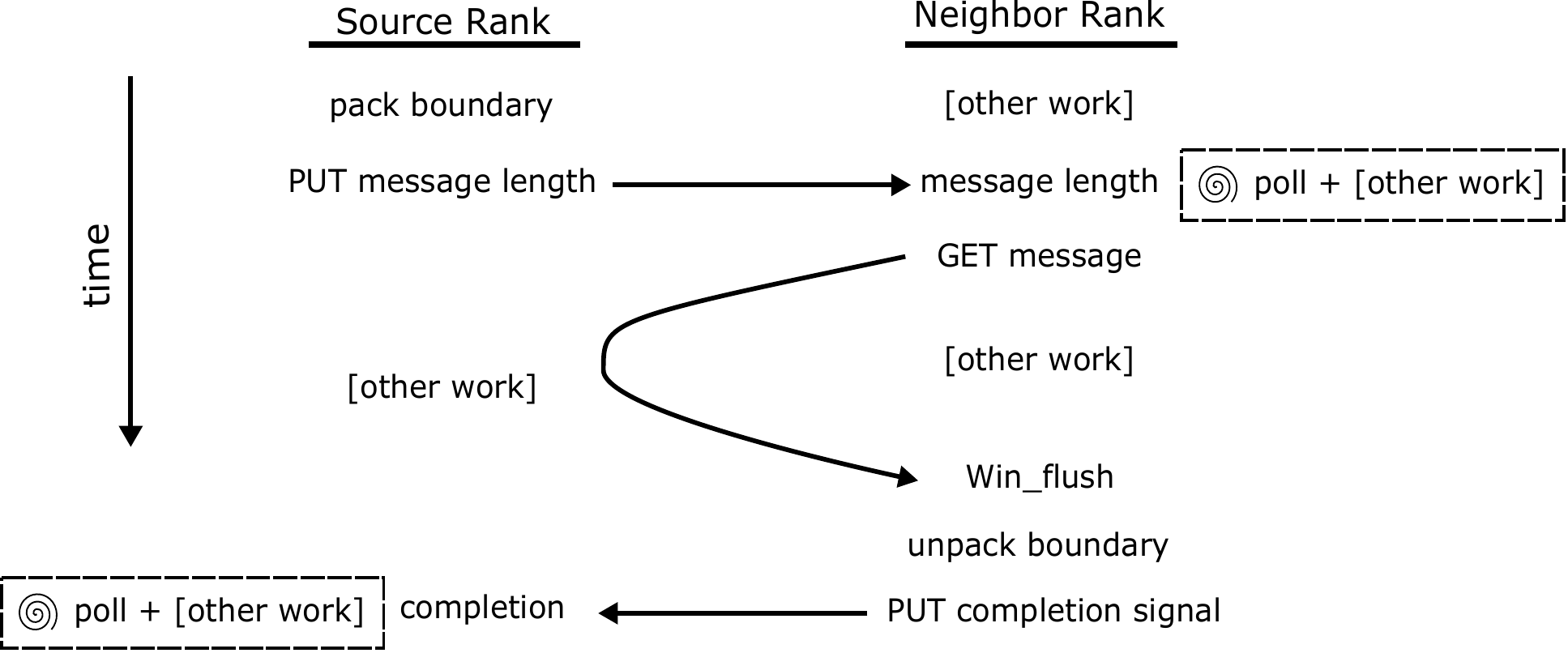}
\caption{MPI-RMA Engine cycle.}
\label{fig:rma}
\end{figure*}

\begin{figure}
	\centering
	\includegraphics[height=.18\textheight]{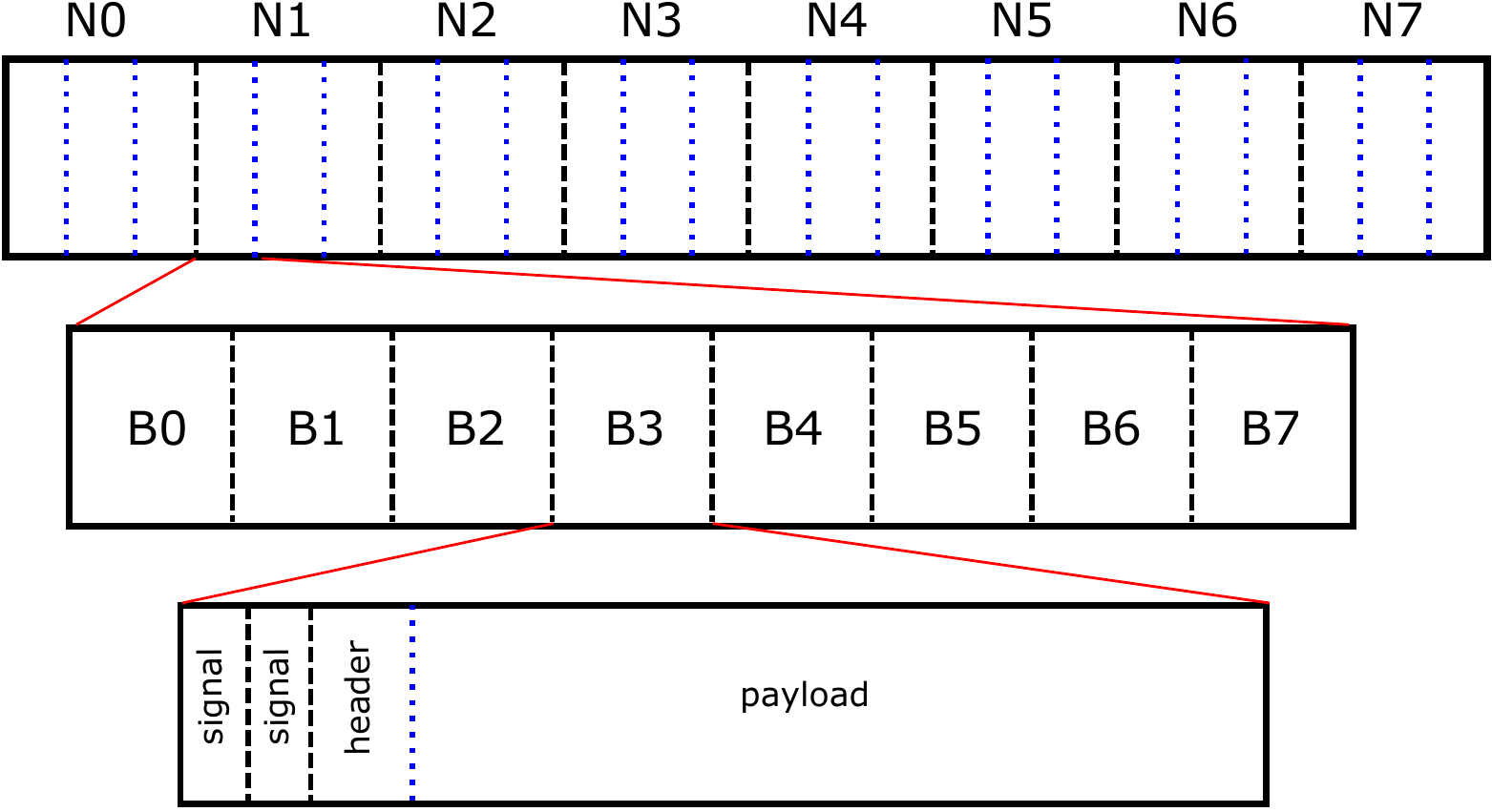}
	\caption{Anatomy of the buffer attached to the RMA window.} \label{fig:rma_anatomy}
\end{figure}

The MPI-RMA Engine handles non-local communication between Patches.  It is generic enough to manage communication of any type of data of a wide range of message lengths with memory overheads and intensity on the network that is run-time tunable. The strategy for the MPI-RMA engine was to remove all explicit synchronization between MPI ranks and utilize all threads for both message packing/unpacking and initiation of network transfers. We use a single passive exposure epoch with MPI-RMA.  The passive epoch starts with ranks calling \textbf{\mbox{MPI\_Win\_lock()}} for each rank it will communicate with.  The communication strategy in WOMBAT does not use protections between ranks, and the lock argument to \textbf{\mbox{MPI\_Win\_lock()}} is always set to \textbf{\mbox{MPI\_LOCK\_SHARED}}.  Locking and unlocking for RMA exposure is moved outside the time loop, which essentially removes their cost in exchange for minimal overhead introduced by a signaling scheme.\par

Figure \ref{fig:rma} shows an overview of the steps in the MPI-RMA Engine.  Operations from the point of view of both a source and neighbor rank are shown in time.  Note that all source ranks are also a neighbor rank, meaning that the steps are symmetric.  The process begins with a source rank packing some (not all) boundary data from a Patch into a buffer.  The rank then sends an 8 Byte signal to the neighbor rank with \textbf{\mbox{MPI\_Put()}} indicating the size of the message that has been packed.  At some point the neighbor rank starts to poll on the local address where this signal is to be deposited waiting for the value to become something other than the initial state.  Reading this address must be done carefully as to not allow the compiler to cache the value in a register.  We do this by performing the read on the signal address from a simple \textbf{C} routine, designed to prevent any register caching from the calling \textbf{Fortran} code.  Once the signal value is modified, the value is interpreted as the message length.  If it is zero there is no message to transfer, which can happen for a variety of reasons due to the generic messaging property of the MPI-RMA Engine.  If the value is greater than zero the neighbor rank initiates a network transfer with an \textbf{\mbox{MPI\_Get()}}.  While the network transfer is in flight, both the source and neighbor rank do other communication or computation work.  At some point later the neighbor rank needs the transfer to complete and calls \textbf{\mbox{MPI\_Win\_flush()}}.  The message is then unpacked, and the neighbor rank then sends a signal pack to the source rank indicating that the transfer is done and the source buffer can be freely modified.  The source rank eventually polls on that signal before it can repeat the full process over again. \par

We note that an alternative implementation of this cycle could be done entirely with \textbf{\mbox{MPI\_Put()}}.  In such a design, a \textbf{\mbox{MPI\_Put()}} call would immediately move data to the destination rank completed sometime later with \textbf{\mbox{MPI\_Win\_flush()}}.  Then the initial \textbf{\mbox{MPI\_Put()}} above is used to signal that data is in the destination buffer.  We did not use this design because it has the potential for generating more intense many-to-one traffic patterns, which can lead to degraded performance on most HPC interconnects.

The MPI-RMA Engine cycle applies to each segment of the single communication buffer that was created for the RMA window with \textbf{\mbox{MPI\_Win\_Allocate()}}.  Multiple segments in this buffer allow for many unique messages to be exchanged with neighbor ranks.  They also present potential thread parallelism for communication. The MPI-RMA Engine cycle is self-contained and can be applied to any number of independent messages to be exchanged with minimal contention or protection required between them.  Multiple threads can therefore drive the engine entirely independently as long as they operate on separate buffer segments. \par

The single buffer attached to the RMA window in WOMBAT is decomposed into multiple regions each available for communicating Patch boundary data.  Figure \ref{fig:rma_anatomy} shows the anatomy of this buffer with an example of a 2d Cartesian domain with 9 MPI ranks (similar to the domain structure in Figure \ref{fig:decomp}).  The figure begins at the top looking at the entire RMA buffer logically separated into equal size segments for each of the 8 neighbors (labeled N0 through N7) any rank might have.  Note that it is possible for some of the logical neighbors to be the same MPI rank if the world grid is periodic.  Each of these segments is further divided based on a run-time tunable value for the number of ``mailboxes'' dedicated to each neighbor rank.  Increasing the number of mailboxes has the effect of putting more network transfers in-flight at any moment, which can reduce the number of iterations in the Engine.  Each of these mailboxes is large enough to buffer all boundary data to and from one Patch (size is doubled for send and receive). It is not necessary or common that this data be from the same source Patch.  In one of these mailboxes, there are 8 boundary segments corresponding to the 4 edges  and 4 corners that will be communicated in 2d from a Patch labeled B0 through B7.  A single section of a boundary segment, there are 4 distinct sections.  The first two are each 8 Bytes in length and are used for the incoming and outgoing signals described above and in Figure \ref{fig:rma}.  The next ``header'' section is used to encode descriptive data about the message payload.  This information includes identifying information for the Patch that should receive this boundary data.  The header can be leveraged for performing other communication that might be useful to exchange between rank on a regular basis, such as load imbalance statistics or changes in the ownership of a given Patch.   The final section in the boundary segment is the message payload.

The MPI-RMA Engine also has methods for initiating and completing non-blocking global reductions.  They are used to compute time step sizes across all MPI ranks.  Our implementation delays time step calculation by one step in order to overlap the collective with work.

\subsubsection{MPI-RMA Thread Optimization in Cray MPICH} \label{mpich_rma}

In the SPMD OpenMP model, threads do their computation, message sending, and message completion asynchronously, so contention on the interconnect resources becomes relevant to performance. On Cray XC systems the Aries interconnect provides 128 hardware ``lanes'' called communication domains (CDMs) for concurrent message transfers and synchronizations (although MPI does not always make use of all of them). The MPI library assigns these CDMs either statically to threads the first time a thread makes an MPI call, or dynamically each time a message is sent or completed. \par
In SPMD OpenMP, static assignment of CDMs to threads is not feasible anymore, because it provides no means for the MPI library to dynamically minimize contention. For example, if a thread needs to complete all messages targeting a specific remote rank, it may need access to several CDMs that have been statically assigned to other threads before.  Safe access to these CDMs could be handled with a mutex, but doing so can force other threads to wait for access to the CDM before sending a message. Hence, dynamic allocation of CDMs is required to minimize overhead from CDM contention and maximize performance. \par
We have adapted Cray's MPI-RMA implementation to use lock-free dynamic allocation of CDMs. It is now designed specifically to minimize overhead due to CDM assignment and maximize performance for SPMD approaches. The library guarantees contention-free communication as long as the number of concurrent requests to send and/or complete a message does not exceed the available number CDMs.


\section{Numerical Methods} \label{methods}

For an initial implementation in the code design discussed above we use a 2$^{nd}$ order, directionally un-split version of the non-relativistic ideal MHD solver described in RJ95 and \citet{1998ApJ...509..244R} referred to as MHDTVD.  This new implementation follows the CTU+CT scheme, described in \citet{2005JCoPh.205..509G} (hereafter GS05) and \citet{2008JCoPh.227.4123G} (hereafter GS08), modified for the MHDTVD solver.  The algorithm outlined here solves the equations of MHD neglecting charge separation between ions and electrons, electrical resistivity, viscosity, and non-adiabatic processes, such as thermal conduction.  With these assumptions the ideal MHD equations are
\begin{align}
	\frac{\partial\rho}{\partial t} + \mathbf{\nabla} \cdot (\rho \mathbf{v}) &= 0, \label{e:mhd_rho} \\
\frac{\partial\mathbf{v}}{\partial t} + \mathbf{v} \cdot \mathbf{\nabla v} + \frac{1}{\rho}\mathbf{\nabla}P - 
\frac{1}{\rho}\left(\mathbf{\nabla \times B}\right) \mathbf{\times B} &= 0, \label{e:mhd_mom} \\
\frac{\partial P}{\partial t} + \mathbf{v} \cdot \mathbf{\nabla}P + \gamma P \mathbf{\nabla} \cdot \mathbf{v} &= 0, \label{e:mhd_energy} \\
\frac{\partial B}{\partial t} - \mathbf{\nabla \times} \left(\mathbf{v \times B}\right) &= 0, \label{e:mhd_field}
\end{align}
where $\gamma$ is the plasma adiabatic index.  Following the convention of RJ95, we have selected our units such that 4$\pi$ does not appear in these equations. For a one-dimensional flow along the $X$ direction, Equations \ref{e:mhd_rho} - \ref{e:mhd_field} can be written in the conservative form
\begin{align}
	\frac{\partial \mathbf{q}}{\partial t} + \frac{\partial \mathbf{F}}{\partial x} &= 0, \label{e:mhd_cons}
\end{align}
where $\mathbf{q}$ and $\mathbf{F}$ are the state vector and flux vector respectively defined as 
\begin{align}
\mathbf{q} &= \left( \begin{array}{c} \rho \\ \rho v_{x} \\ \rho v_{y} \\ \rho v_{z} \\ B_{x} \\ B_{y} \\ B_{z} \\ E \end{array} \right), \\
	\mathbf{F} &= \left( \begin{array}{c}
                                     \rho v_{x} \\
                                     \rho v_{x}^{2} + P^{*} - B_{x}^{2} \\
                                     \rho v_{x}v_{y} - B_{x}B_{y} \\
                                     \rho v_{x}v_{z} - B_{x}B_{z} \\
                                     0 \\
                                     B_{y}v_{x} - B_{x}v_{y} \\
                                     B_{z}v_{x} - B_{x}v_{z} \\
                                     (E + P^{*})v_{x} - B_{x}(B_{x}v_{x} + B_{y}v_{y} + B_{z}v_{z})
                                     \end{array} \right).
\end{align}
The total pressure and total energy are given by
\begin{align}
	P^{*}& = P + \frac{1}{2}\left(B_{x}^{2} + B_{y}^{2} + B_{z}^{2}\right) \label{e:totpressure} \\
	E &= \frac{P}{\gamma - 1} + \frac{\rho}{2}\left(v_{x}^{2} + v_{y}^{2} + v_{z}^{2}\right) + \frac{1}{2}\left(B_{x}^{2} + B_{y}^{2} + B_{z}^{2}\right). \label{e:totenergy}
\end{align}
A source term vector can be added to Equation \ref{e:mhd_cons} to include additional physics, such as gravity, geometry corrections, cooling, and cosmic-ray feedback.  This system of equations is hyperbolic under the definition that the Jacobian matrix, $\mathbf{A} = \partial \mathbf{F} / \partial \mathbf{q}$, has all real eigenvalues and a complete set of right eigenvectors.  This system is not strictly hyperbolic, however, due to conditions that can produce degenerate eigenvalues.  The seven eigenvalues $a_{1,7} = v_{x} \pm c_{f}$, $a_{2,6} = v_{x} \pm c_{a}$, $a_{3,5} = v_{x} \pm c_{s}$, and $a_{4} = v_{x}$ correspond to three MHD wave families and an entropy mode.  The characteristic wave speeds are
\begin{align}
	c_{f} &= \left(\frac{1}{2} \left[a^{2} + \frac{B_{x}^{2} + B_{y}^{2} + B_{z}^{2}}{\rho} + \right. \right. \nonumber \\
	   & \left. \left. \sqrt{\left(a^{2} + \frac{B_{x}^{2} + B_{y}^{2} + B_{z}^{2}}{\rho}\right)^2 -
        4a^{2}\frac{B_{x}^{2}}{\rho}} \right] \right)^{1/2} \\
		c_{s}& = \left(\frac{1}{2} \left[a^{2} + \frac{B_{x}^{2} + B_{y}^{2} + B_{z}^{2}}{\rho} - \right. \right. \nonumber \\
	   & \left. \left. \sqrt{\left(a^{2} + \frac{B_{x}^{2} + B_{y}^{2} + B_{z}^{2}}{\rho}\right)^2 -
        4a^{2}\frac{B_{x}^{2}}{\rho}} \right] \right)^{1/2} \\
		c_{a}& = \sqrt{\frac{B_{x}^{2}}{\rho}},
\end{align}
where the sound speed is defined as $a = \sqrt{\gamma P / \rho}$.  One of the difficulties in solving Equation \ref{e:mhd_cons} is that some of the eigenvalues will coincide in limiting cases and special care must be taken to avoid singularities around points where $B_{x} = 0$ or $B_{y} = B_{z} = 0$ (RJ95).  We summarize the one dimensional MHDTVD algorithm in \S \ref{tvd}.

\subsection{MHD in Two Dimensions}

The 2d directionally un-split update closely follows the steps for the CTU+CT scheme described in GS05.  Our implementation utilizes 5 boundary zones, which requires only one boundary exchange per time step for both state variables and zone corner EMFs.  Given a time step $\Delta t$, the steps in the algorithm are:

\begin{description}
\item [Step 1] Compute the directionally split fluxes in both X and Y directions using initial states $\mathbf{q}^{n}$ from Equation \ref{e:mhdtvd_flux2} for a time step $\Delta t$.
\item [Step 2] Compute a zone-centered reference EMF for use in the mid-time step constrained transport update of the face-centered magnetic field.  The EMF is given by $v_x \times B_y + v_y \times B_x$ with each input derived from the initial state vector $\mathbf{q}^{n}$.
\item [Step 3] Using the upwinded algorithm in GS05, compute EMF values at zone corners using the B$_y$ and B$_x$ fluxes from the X and Y passes from Step 1 and the reference EMF from Step 2.
\item [Step 4] Update the face centered magnetic field $\mathbf{b}^{n}$ to $\mathbf{b}^{n+1/2}$  from the EMFs in Step 3 over $\Delta t / 2$.
\item [Step 5] Update the zone centered state vector from the initial states $\mathbf{q}^{n}$ to $\mathbf{q}^{n+1/2}_x$ using fluxes from the Y pass in Step 1 applied over $\Delta t / 2$.  Include the $\nabla \cdot B$ source term vector described by GS05.
\item [Step 6] Using the preconditioned state $\mathbf{q}^{n+1/2}_x$, compute fluxes along X from Equation \ref{e:mhdtvd_flux2} for a time step $\Delta t$.
\item [Step 7] Repeat steps 5 and 6 for the Y direction.
\item [Step 8] Compute a zone-centered reference EMF for use in the final CT update of the face-centered magnetic field.  The EMF is given by $v_x \times B_y + v_y \times B_x$ with $v_x$ and $v_y$ coming from an un-split update of $\mathbf{q}^{n}$ to $\mathbf{q}^{n+1/2}$ using the fluxes from Steps 6 and 7.
\item [Step 9] Using the upwinded algorithm in GS05, compute EMF values at zone corners using the B$_y$ and B$_x$ fluxes from the X and Y passes from Steps 6 and 7 and the reference EMF from Step 8.
\item [Step 10] Use an un-split update of the state vector $\mathbf{q}^{n}$ to $\mathbf{q}^{n+1}$ using fluxes from Steps 6 and 7 applied over $\Delta t$.
\item [Step 11] Update the face centered magnetic field $\mathbf{b}^{n}$ to $\mathbf{b}^{n+1}$ from the EMFs in Step 9 over $\Delta t$.  Update the zone centered magnetic field from averages of the face centered magnetic field as described in GS05.
\end{description}

\subsection{MHD in Three Dimensions}

The 3d un-split update is based on the 6-solve algorithm described in GS08.  We again utilize 5 boundary zones as described above for 2d.  The steps in the 3d algorithm are:

\begin{description}
\item [Step 1] Compute the directionally split fluxes in the X, Y and Z directions using initial states $\mathbf{q}^{n}$ from Equation \ref{e:mhdtvd_flux2} for a time step $\Delta t$.
\item [Step 2] Compute zone-centered reference EMFs for use in the mid-time step constrained transport update of the face-centered magnetic field using inputs derived from the initial state vector $\mathbf{q}^{n}$.
\item [Step 3] Using the upwinded algorithm in GS08, compute EMF values at zone corners using the magnetic fluxes from the directional passes from Step 1 and the reference EMF from Step 2.
\item [Step 4] Update the face centered magnetic field $\mathbf{b}^{n}$ to $\mathbf{b}^{n+1/2}$  from the EMFs in Step 3 over $\Delta t / 2$.
\item [Step 5] Update the zone centered state vector from the initial states $\mathbf{q}^{n}$ to $\mathbf{q}^{n+1/2}_x$ with an un-split update using fluxes from the Y and Z passes in Step 1 applied over $\Delta t / 2$.  Include the $\nabla \cdot B$ source term vector described by GS08.
\item [Step 6] Using the preconditioned state $\mathbf{q}^{n+1/2}_x$, compute fluxes along X from Equation \ref{e:mhdtvd_flux2} for a time step $\Delta t$.
\item [Step 7] Repeat steps 5 and 6 for the Y and Z directions using the appropriate transverse fluxes from Step 1.
\item [Step 8] Compute zone-centered reference EMFs for use in the final CT update of the face-centered magnetic field.  Velocity values come from an un-split update of $\mathbf{q}^{n}$ to $\mathbf{q}^{n+1/2}$ using the fluxes from Steps 6 and 7.
\item [Step 9] Using the upwinded algorithm in GS08, compute EMF values at zone corners using the magnetic fluxes from the directional passes from Steps 6 and 7 and the reference EMF from Step 8.
\item [Step 10] Use an un-split update of the state vector $\mathbf{q}^{n}$ to $\mathbf{q}^{n+1}$ using fluxes from Steps 6 and 7 applied over $\Delta t$.
\item [Step 11] Update the face centered magnetic field $\mathbf{b}^{n}$ to $\mathbf{b}^{n+1}$ from the EMFs in Step 9 over $\Delta t$.  Update the zone centered magnetic field from averages of the face centered magnetic field as described in GS08.
\end{description}


\section{Test Calculations} \label{tests}

\subsection{Linear Wave Convergence}
\label{wave_conv}
We performed linear wave convergence tests using eigenvectors of the Roe matrices for hydrodynamics and MHD following the setup used by GS05. For one-dimensional tests, we use a periodic domain $L = 1$ divided into $N$ zones containing a background fluid with $\rho = 1$, $P = 3/5$, and $\gamma = 5/3$. The background is at rest for shear and entropy waves, otherwise $v_x = 1$. For hydrodynamic waves (sound, $v_y$ and $v_z$ shear, and entropy modes), $B_x = B_y = B_z = 0$, while the background for MHD waves (slow, Alfv\'{e}n, fast, and entropy modes) has magnetic field components $B_x = 1, B_y = \sqrt{2}, B_z = 1/2$. A sinusoidal perturbation is applied to this background state, such that the initial state vector is given by $\mathbf{q}_0 = \bar{\mathbf{q}} + A_0 \mathbf{R}_k \cos(2\pi x)$, where $\bar{\mathbf{q}}$ is the background state, $A_0 = 10^{-6}$ is the amplitude, and $R_k$ is the right eigenvector for the wave mode $k$.

\begin{figure}
\includegraphics[height=.25\textheight]{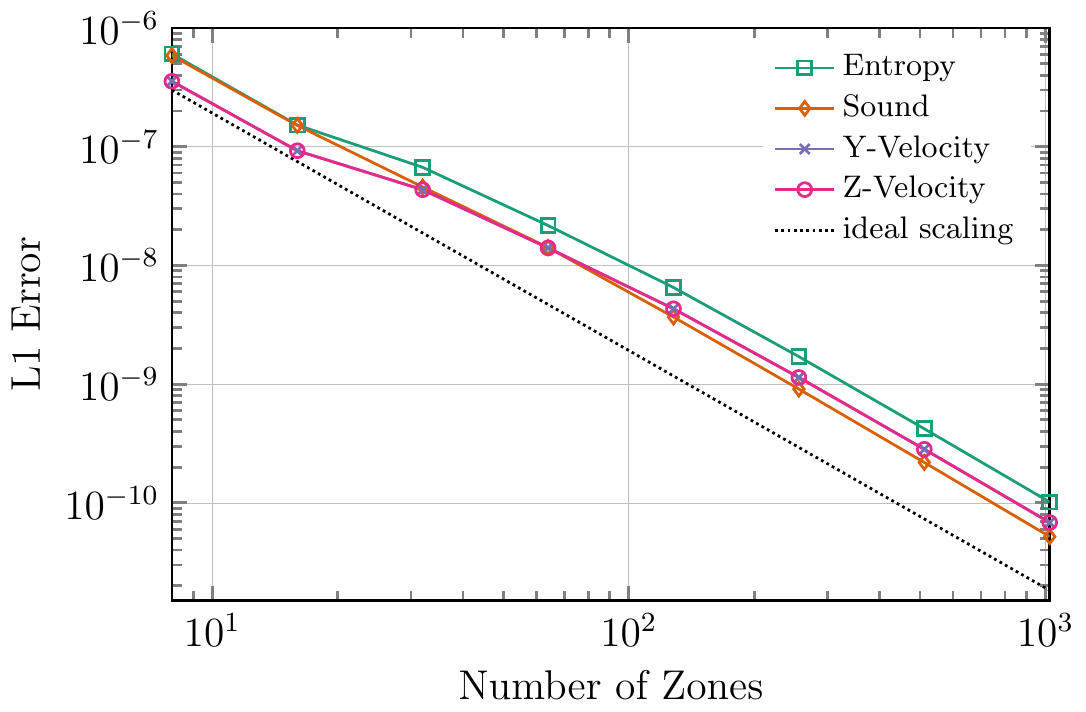}
\includegraphics[height=.25\textheight]{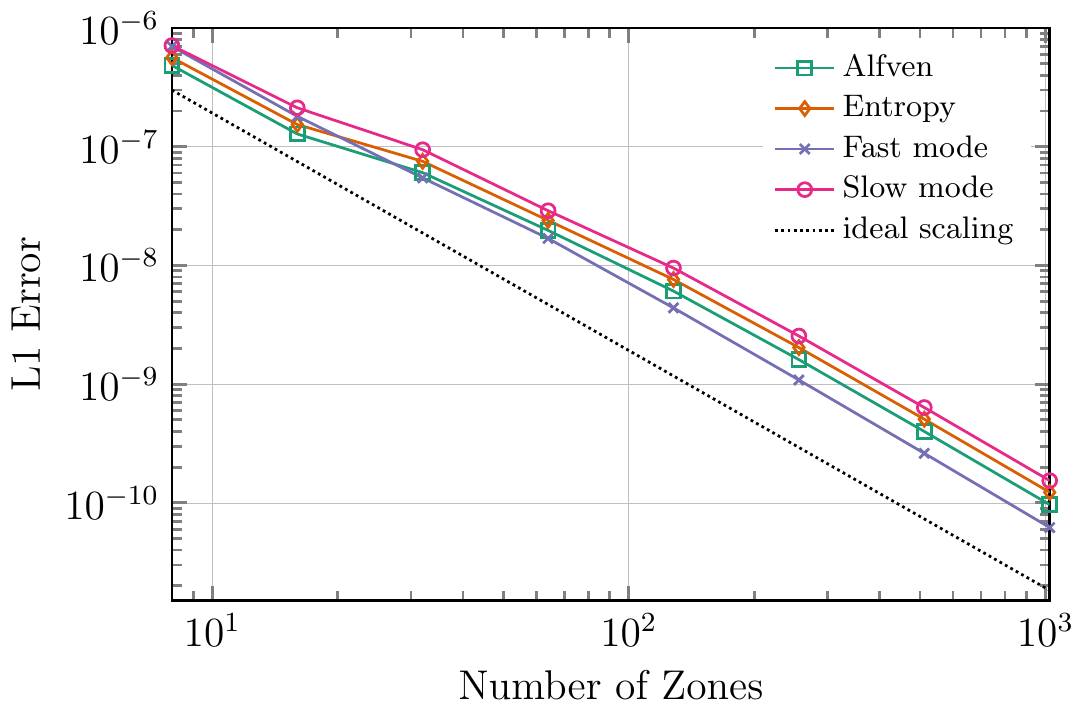}
\includegraphics[height=.25\textheight]{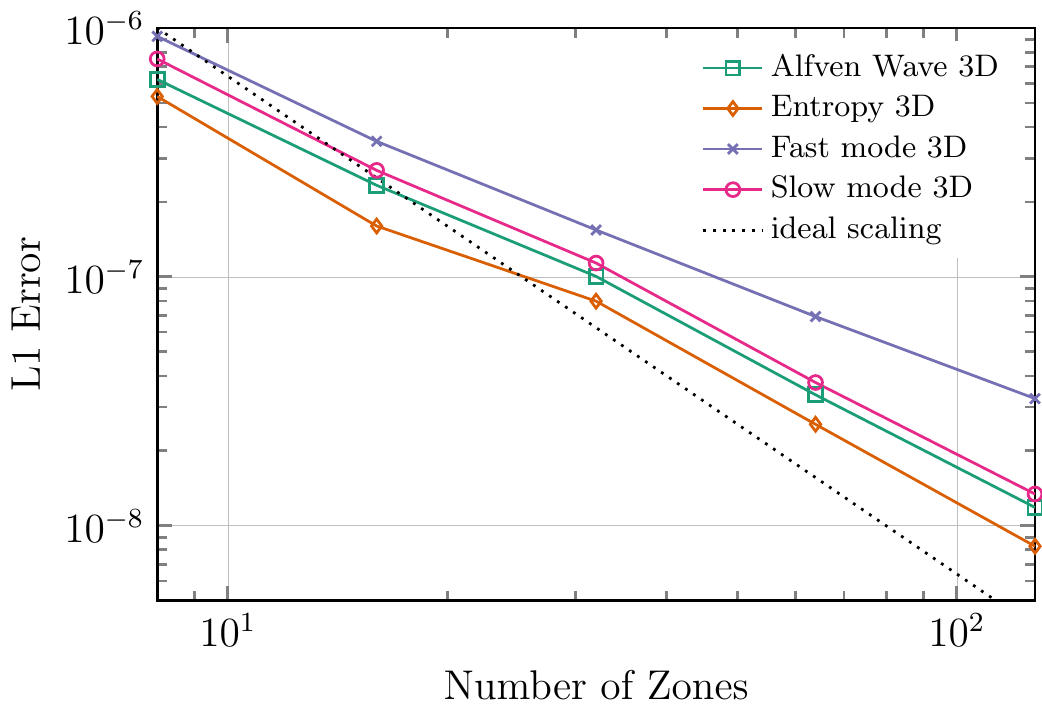}
\caption{Convergence of the norm of the $L_1$ error vector for wave modes after propagating one wavelength. The dotted line shows a slope of -2 for comparison.  The \emph{top} plot shows hydrodynamic wave modes in 1d.  The \emph{middle} plot shows MHD wave modes in 1d.  The \emph{bottom} plot shows MHD wave modes in 3d oblique to the grid. }
\label{fig:wave_conv}
\end{figure}

Each wave is propagated for one wavelength, and then the error in the solution is computed using the $L_1$ error vector averaged over every zone $i$, defined by $\delta \mathbf{q} = N^{-1} \sum\limits_{i}|\mathbf{q}_i - \mathbf{q}_{i,0}|$. Increasing the number of zones up to $N = 1024$, the solution for each wave mode in 1d converges with second-order accuracy as seen by the norm of the $L_1$ error vector in Figure \ref{fig:wave_conv}.

We also tested the convergence of MHD waves propagating oblique to a three-dimensional grid, following the setup in GS08. The wave is initialized rotated with respect to a computational grid of size $(L_X, L_Y, L_Z) = (3, 3/2, 3/2)$ with 2NxNxN zones, such that the wave vector is $\vec{k} = (1/3, 2/3, 2/3)$. The face centered magnetic field components are initialized via a vector potential defined at the corners of the grid zones, and then zone centered magnetic field values are averaged from face centered fields. After propagating one wavelength, the $L_1$ error vector is computed with respect to the initial conditions. The convergence with increasing resolution is shown in Figure \ref{fig:wave_conv}.

\subsection{RJ95 2a}
\begin{figure*}
\centering
\includegraphics[height=.8\textheight]{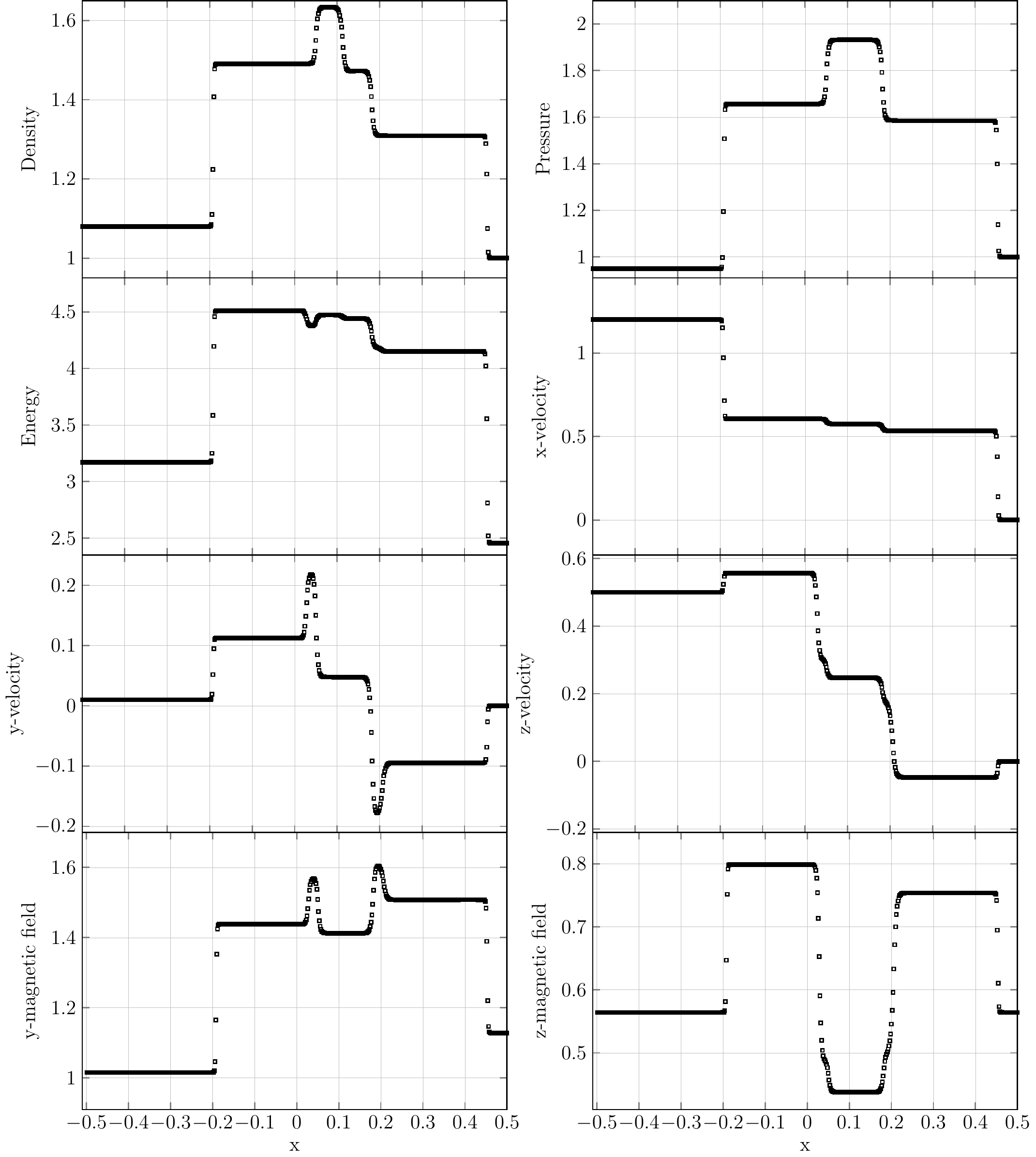}
\caption{RJ95 shock tube test 2a in 1d with 512 zones at t = 0.2. Slices of density, pressure, energy, velocity components, and magnetic field components are shown from top left to bottom right.}
\label{fig:RJ2a}

\end{figure*}

\begin{figure*}
\centering
\includegraphics[height=.8\textheight]{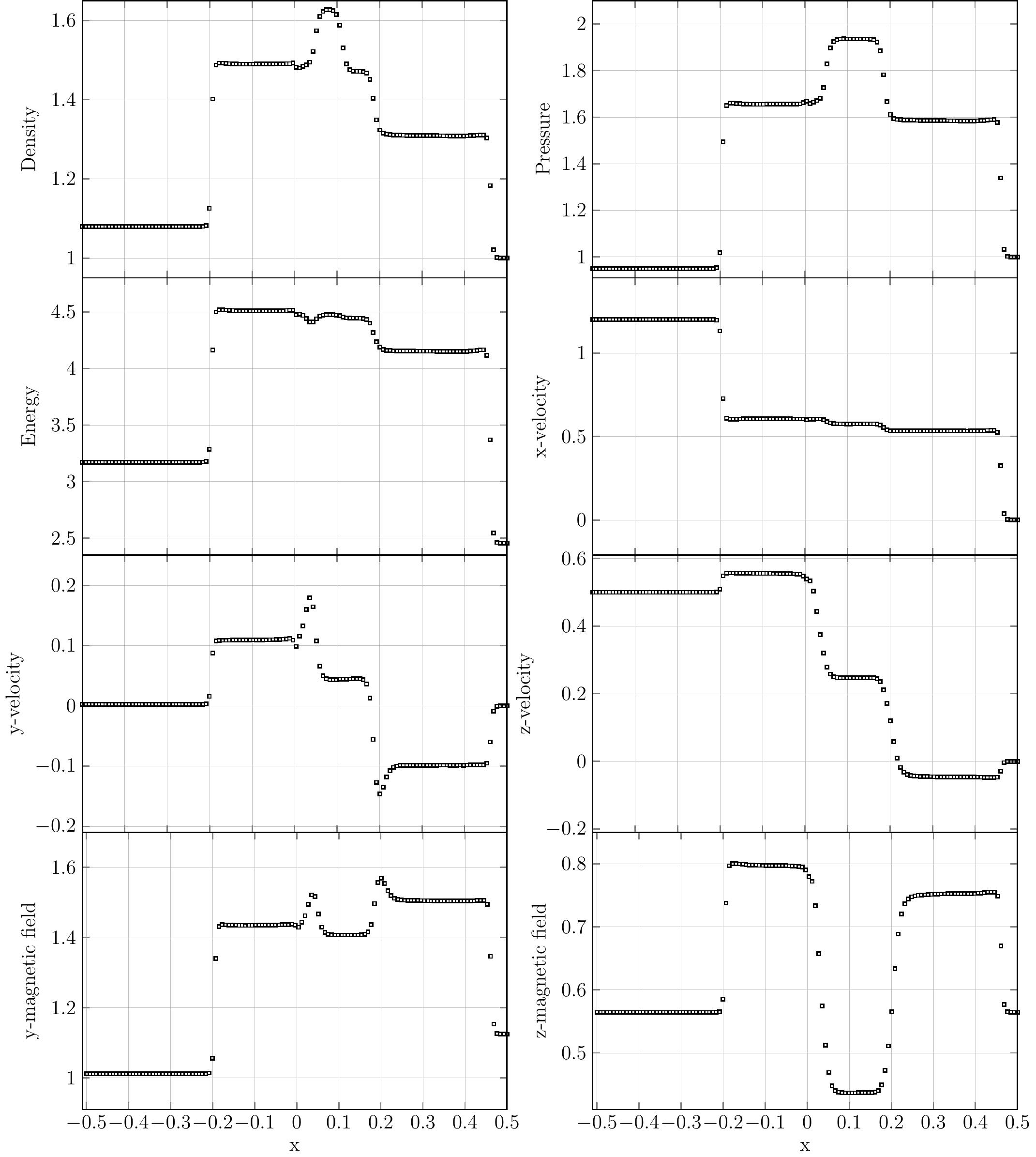}
\caption{RJ95 shock tube test 2a in 3d with 128 zones at t = 0.2. Slices of density, pressure, energy, velocity components, and magnetic field components are shown oblique to the grid from top left to bottom right.}
\label{fig:RJ2a3d}
\end{figure*}

The next set of tests are of the shock-tube setup 2a from RJ95.  The left-hand state was initialized with $(\rho,v_{x},v_{y},v_{z},B_{y},B_{z},P)$ = $[1.08,1.2,0.01,$\allowbreak$3.6/(4\pi)^{1/2},2/(4\pi)^{1/2},0.95]$, and the right-hand state with $[1,0,0,0,4/(4\pi)^{1/2},2/(4\pi)^{1/2},1]$.  For this test $B_{x} = 2/(4\pi)^{1/2}$.  Figures \ref{fig:RJ2a} and \ref{fig:RJ2a3d} shows the evolved grid at $t$ = 0.2 in 1d and 3d respectively.  The shock normal is rotated 45$^\circ$ out of all primary planes in 3d.

\subsection{RJ95 4a}

\begin{figure*}
\centering
\includegraphics[height=.8\textheight]{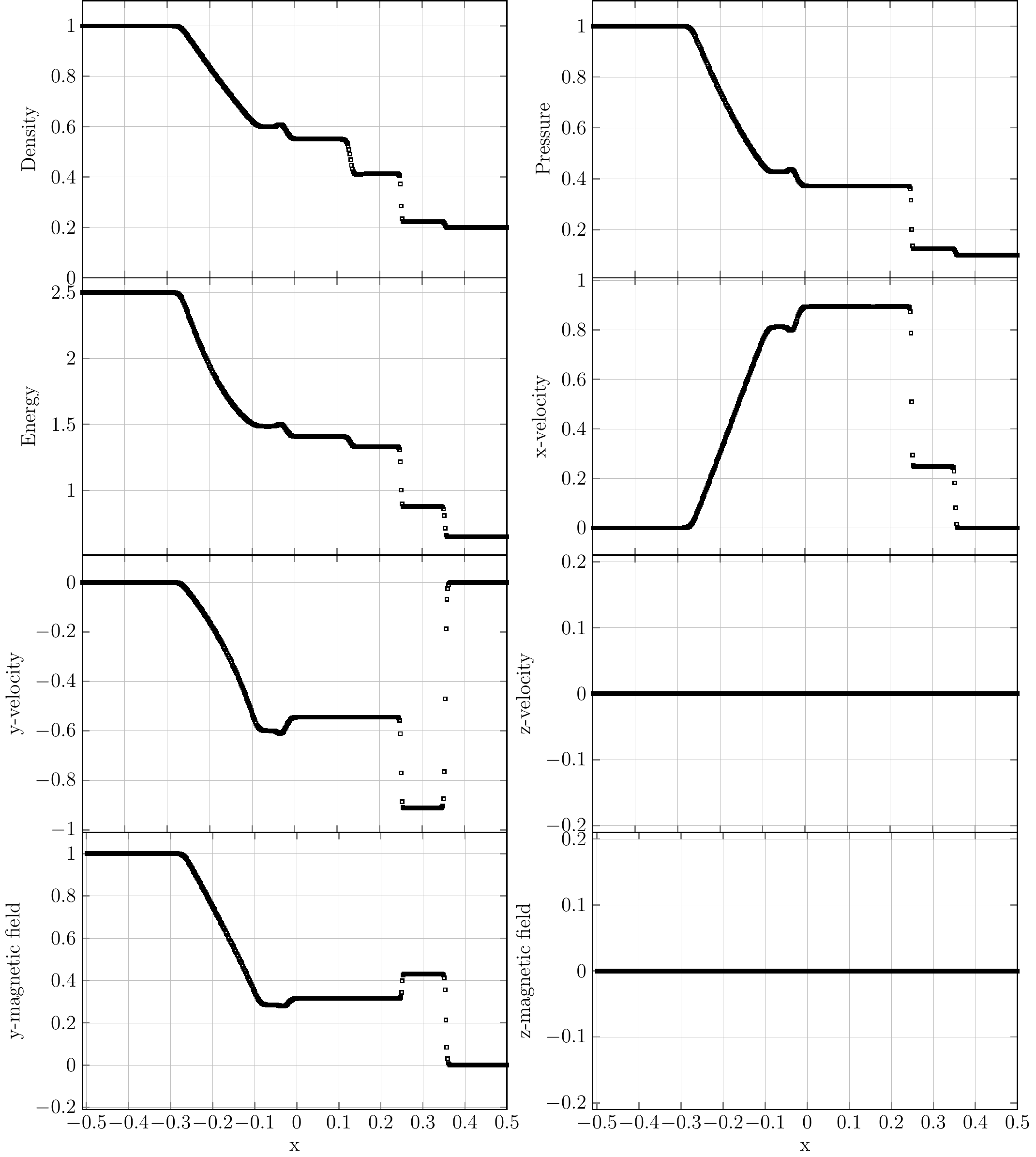}
\caption{RJ95 shock tube test 4a in 1d with 512 zones at t = 0.15. Density, pressure, energy, velocity components, and magnetic field components are shown from top left to bottom right.}
\label{fig:RJ4a}
\end{figure*}

\begin{figure*}
\centering
\includegraphics[height=.8\textheight]{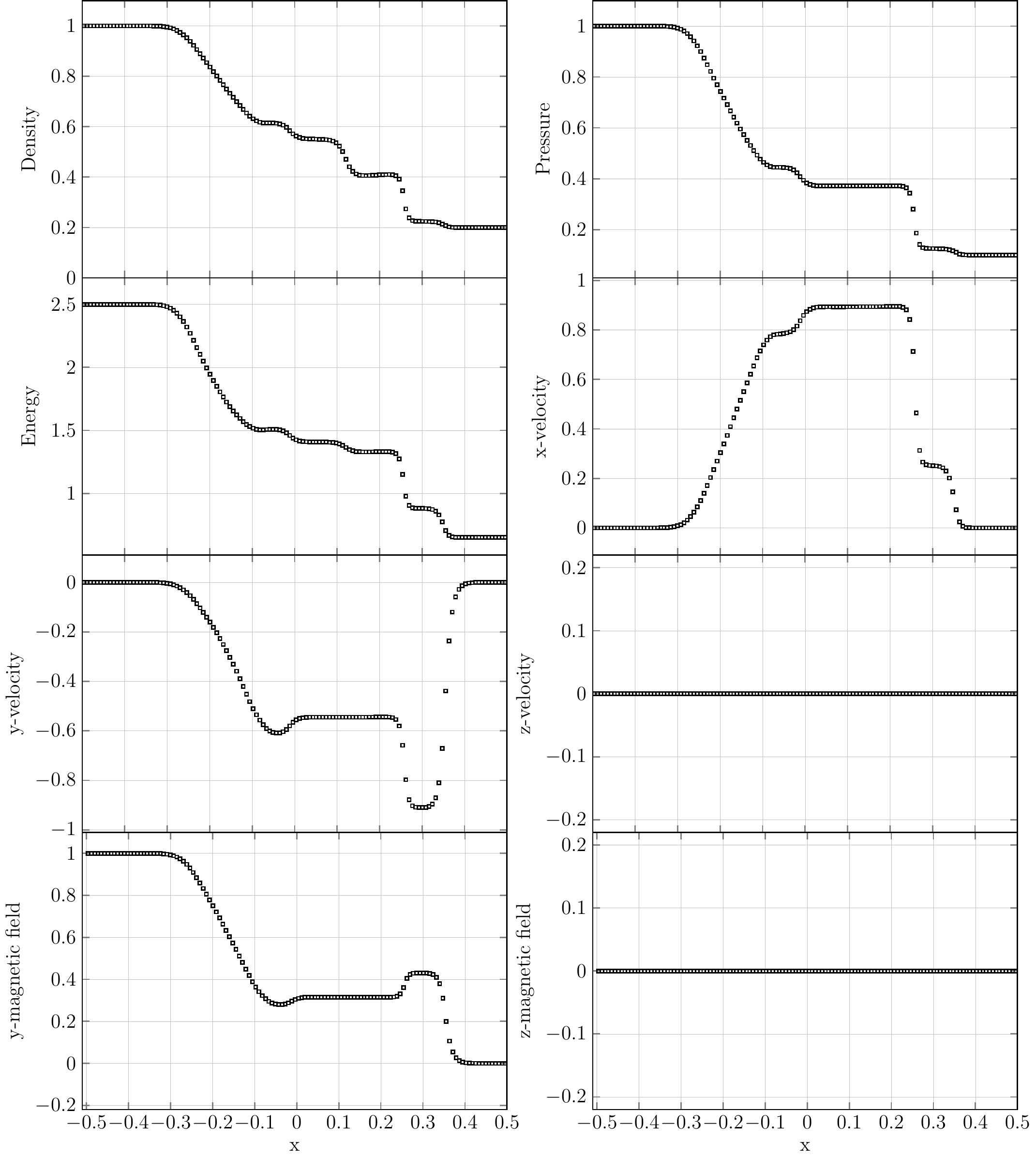}
\caption{RJ95 shock tube test 4a in 3d with 128 zones at t = 0.15. Density, pressure, energy, velocity components, and magnetic field components are shown oblique to the grid from top left to bottom right.}
\label{fig:RJ4a3d}
\end{figure*}

Figures \ref{fig:RJ4a} and \ref{fig:RJ4a3d} show the results at $t$ = 0.15 of the 4a setup from RJ95.  Figure \ref{fig:RJ4a} is the 1d result, and Figure \ref{fig:RJ4a3d} is the 3d result with the shock normal rotated 45$^\circ$ out of all primary planes.  The left-hand state was initialized with $(\rho,v_{x},v_{y},v_{z},B_{y},B_{z},P)$ = $[1,0,0,0,1,0,1]$, and the right-hand state with $[0.2,0,0,0,0,0,0.1]$.  For this test $B_{x} = 1$.

\subsection{Brio \& Wu Shock-tube}
\label{bw}
\begin{figure*}
\centering
\includegraphics[height=.65\textheight]{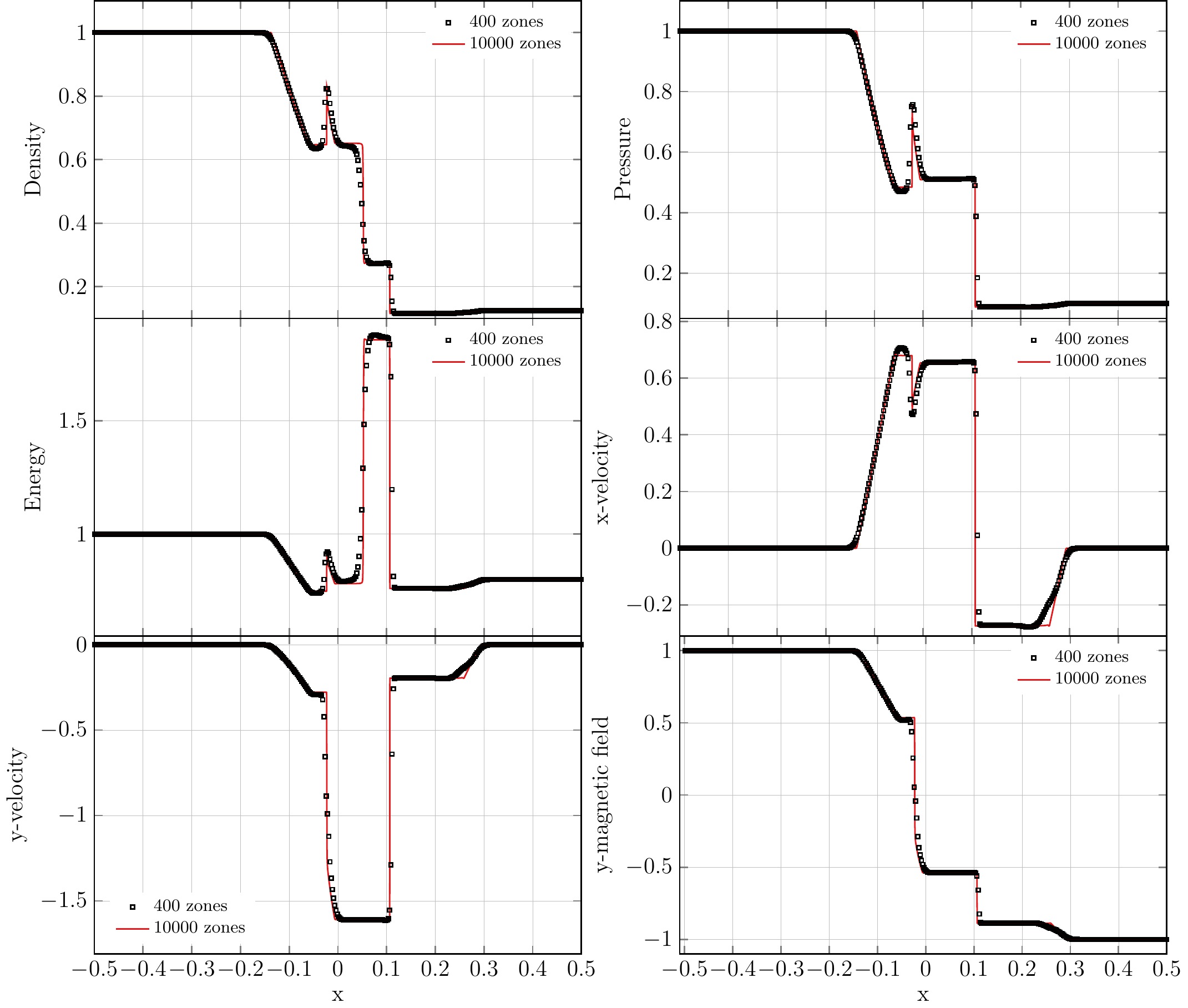}
\caption{Brio \& Wu shock tube test at t = 0.08. The plot for energy is derived from the ratio of pressure to density. The points represent the solution computed on a 400 zone domain, while the solid line is the solution computed on a $10^4$ zone domain.}
\label{fig:1d_brio_wu}
\end{figure*}

We performed the well-known MHD shock tube test of \citet{1988JCoPh..75..400B} on a one-dimensional domain. The left-hand state is initialized with the state vector $(\rho,v_{x},v_{y},v_{z},B_{y},B_{z},P) = (1.0, 0, 0, 0, 1.0, 0, 1.0)$, while the initial right-hand state is $(\rho,v_{x},v_{y},v_{z},B_{y},B_{z},P) = (0.125, 0, 0, 0, -1.0, 0, 0.1)$. Throughout the domain, $B_x = 0.75$, while the adiabatic index $\gamma = 2$. The solution computed on a domain with 400 zones at $t = 0.08$ compared to a better converged solution computed with $10^4$ zones is shown in figure \ref{fig:1d_brio_wu}.

\subsection{Orszag-Tang Vortex}
\label{ot}
\begin{figure*}
\centering
\includegraphics[height=.4\textheight]{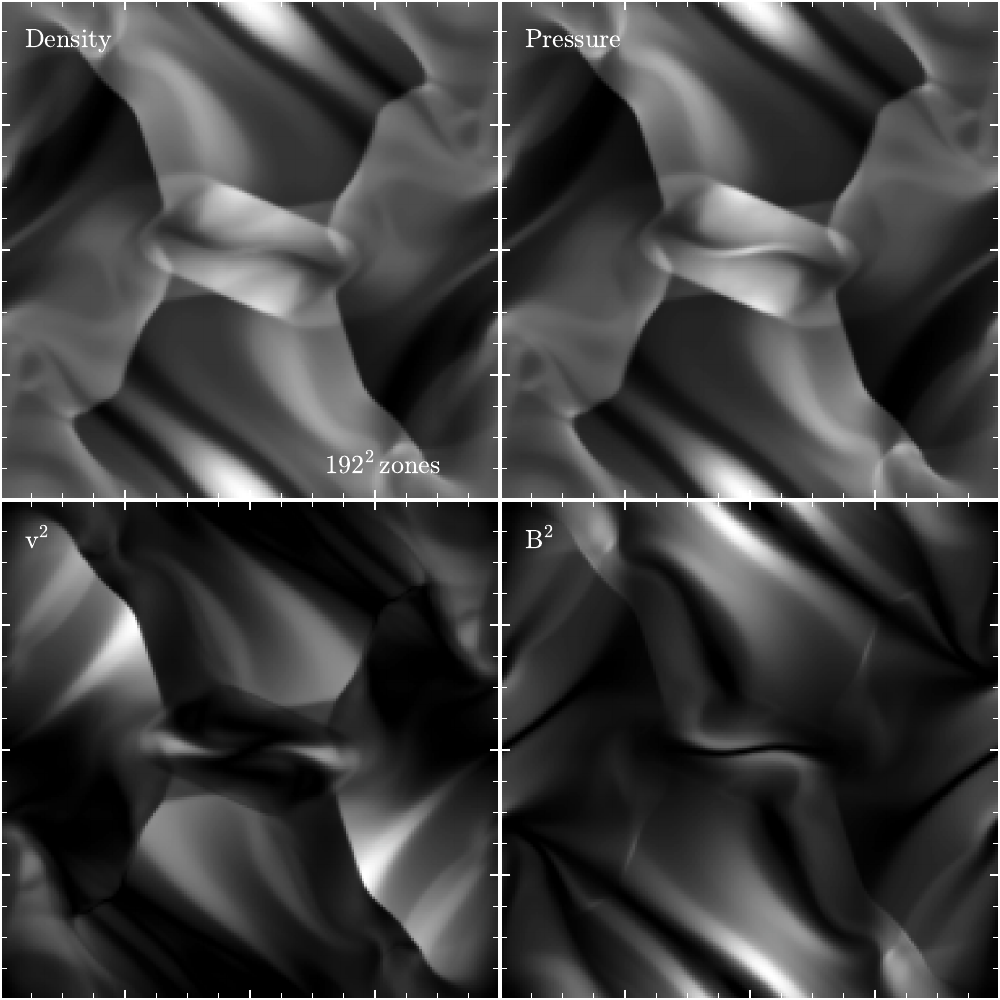}
\includegraphics[height=.4\textheight]{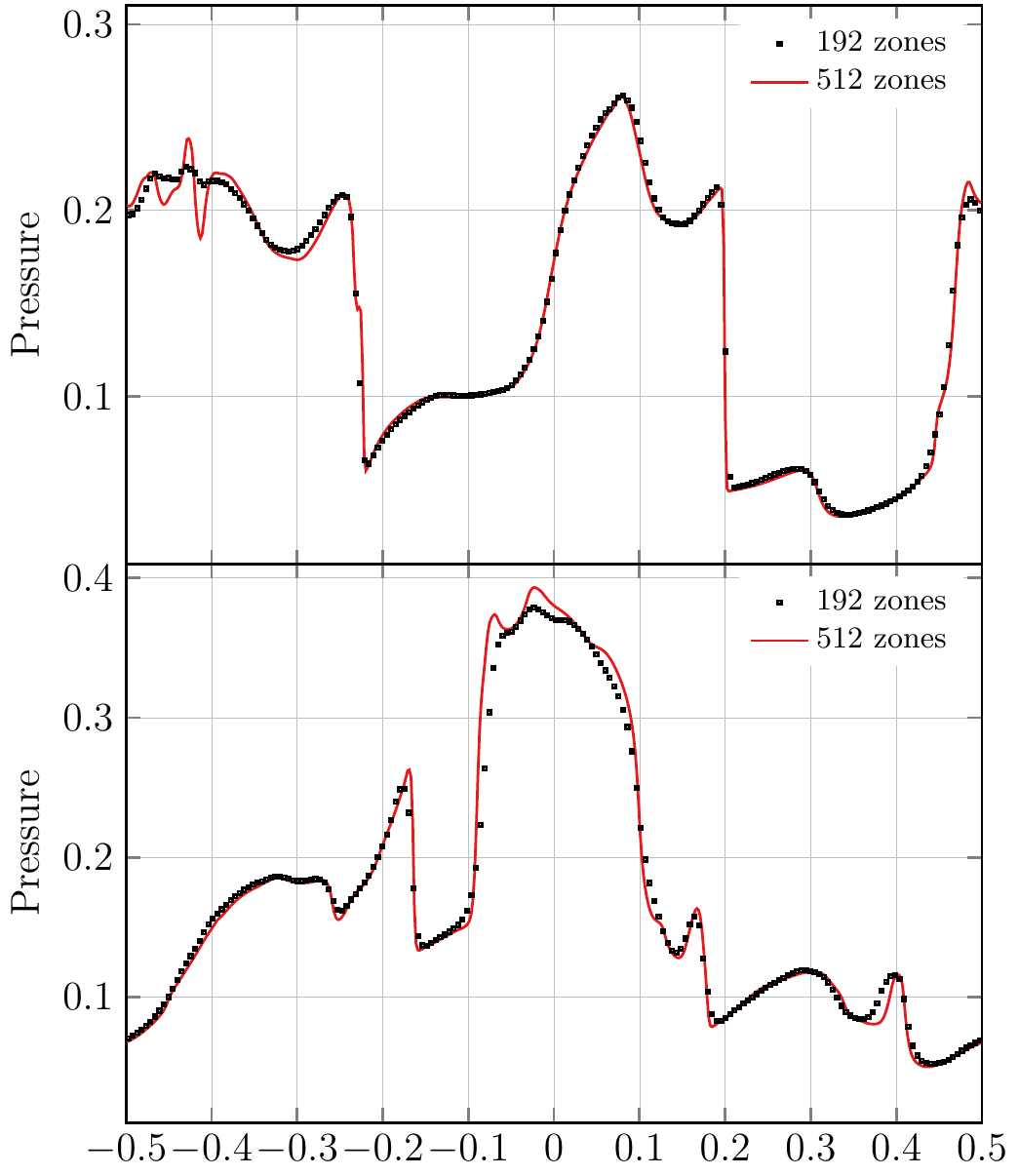}
\caption{Images of selected quantities (\emph{left}) and slices of pressure (\emph{right}) for the Orszag-Tang vortex test at t = 0.5.  Each quantity is scaled black to white linearly from the minimum value to maximum value for a solution computed on a 192x192 zone domain. The slices are at y = -0.1875 (\emph{top}) and y = - 0.073 (\emph{bottom}) and compare the solution computed with 192x192 zones to a solution computed on a domain with 512x512 zones (red lines).} \label{fig:OT}
\end{figure*}

A very common test in 2d for an MHD code is the compressible Orszag-Tang vortex.  This problem was first studied by \cite{1979JFM....90..129O} and is now used as a standard comparison of MHD codes \citep{2008ApJS..178..137S}.  The setup for this problem uses a periodic box with $L_{X}$ = [-0.5,0.5] and $L_{Y}$ = [-0.5,0.5] and 192x192 zones.  Uniform density and pressure are initialized throughout the grid with $\rho = 25/36\pi$, $P = 5/12\pi$ and $\gamma = 5/3$, giving a sound speed of $c_{s} = 1$.  The velocity was initialized as $v_{x} = -v_{0}\text{SIN}(2\pi y)$ and $v_{y} = v_{0}\text{SIN}(2\pi x)$, where $v_{0} = 1$.  The magnetic field along zone faces was derived from the vector potential defined at zone corners $A_{z} = B_{0}\left[\text{COS}(4\pi x)/2 + \text{COS}(2\pi y)\right]/2\pi$, where $B_{0} = 1 / (4\pi)^{1/2}$, with $\mathbf{b} = \nabla \times \mathbf{A}$. Figure \ref{fig:OT} shows the resulting density, gas pressure, specific kinetic energy, and magnetic pressure at $t = 0.5$, as well as slices of the gas pressure at $y = -0.0723$ and $y = -0.1875$.

\subsection{MHD Rotor}

\begin{figure*}
\centering
\includegraphics[height=.4\textheight]{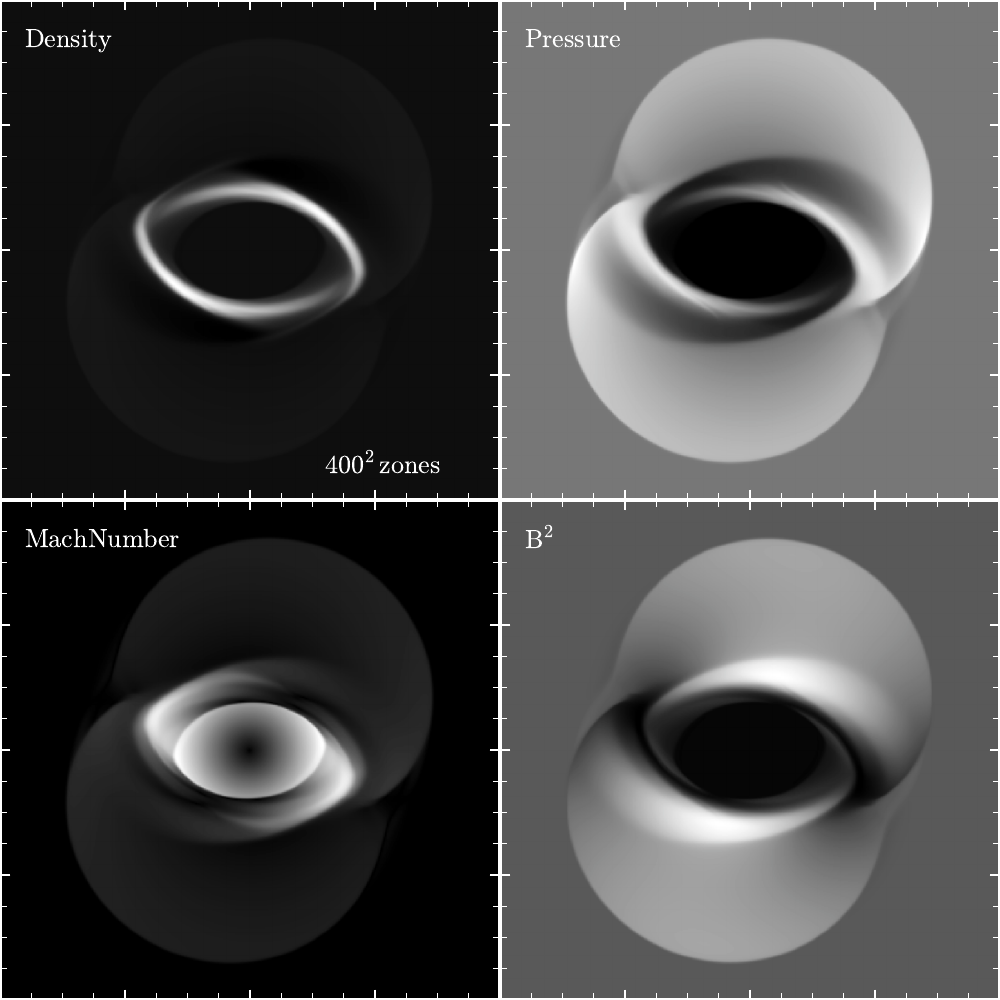}
\includegraphics[height=.4\textheight]{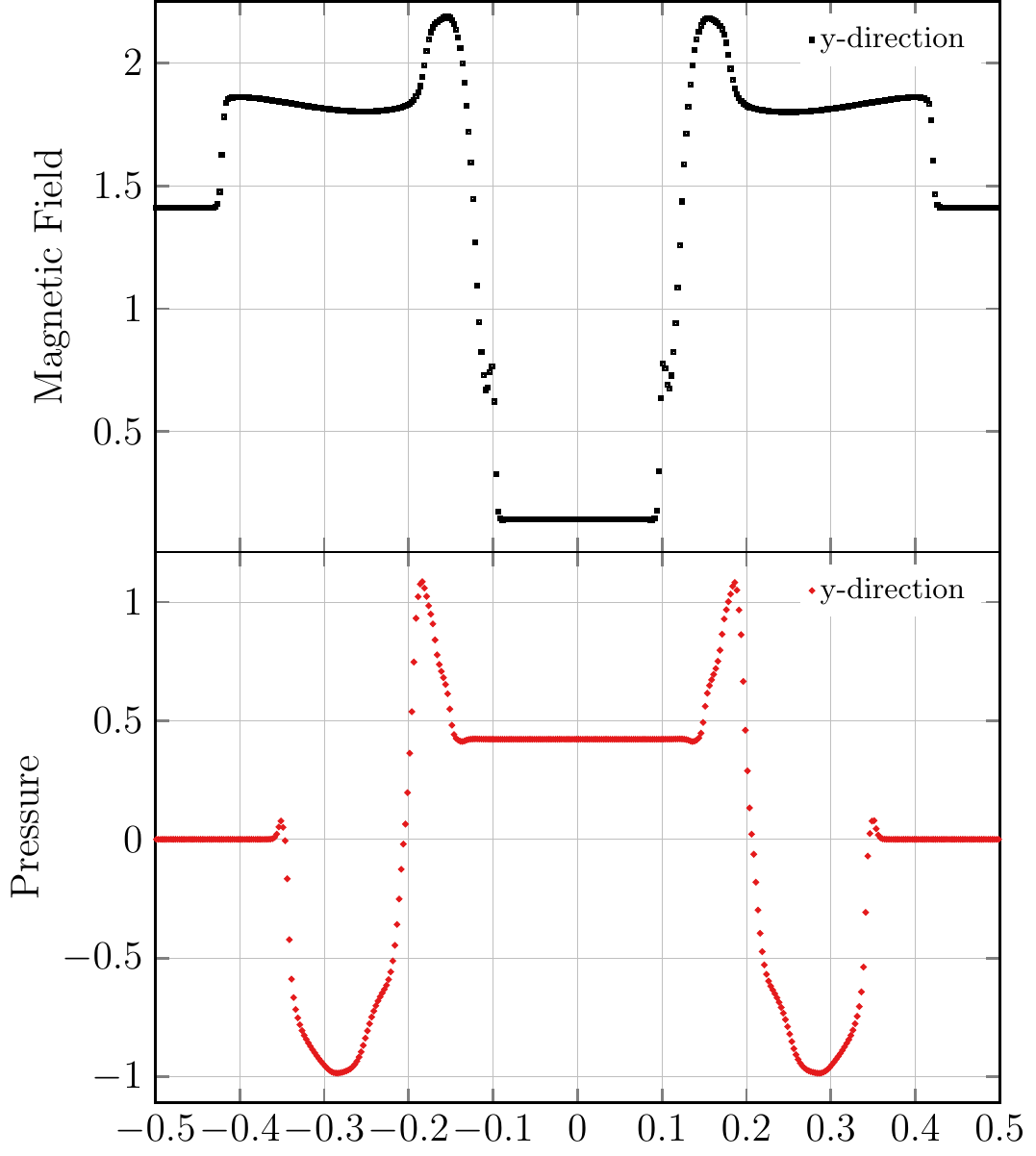}
\caption{Images of selected quantities (\emph{left}) and slices of magnetic field components (\emph{right}) for the MHD rotor test at t = 0.15. Each quantity is scaled black to white from the maximum to minimum value for a solution computed on a 400x400 zone domain. The top slice is taken at y = 0 and shows the y-component of the magnetic field, and the bottom slice is taken at x = 0 and shows the x-component of the magnetic field.}
\label{fig:Rotor}
\end{figure*}

Another common MHD test problem in 2d is that of a rotating disk in a magnetized medium \citep{1999JCoPh.149..270B}. We follow the setup used by \citet{2008ApJS..178..137S} and defined in \citet{2000JCoPh.161..605T} as ``Rotor Problem 1'' on a periodic domain with 400x400 zones. Distributions of density, pressure, Mach number, and magnetic pressure for the solution at $t = 0.15$ is shown in figure \ref{fig:Rotor}, along with slices of the y-component of the magnetic field at $y = 0$ and the x-component of the magnetic field at $x = 0$.

\subsection{Advection of a Field Loop}
\begin{figure*}
\centering
\includegraphics[height=.3\textheight]{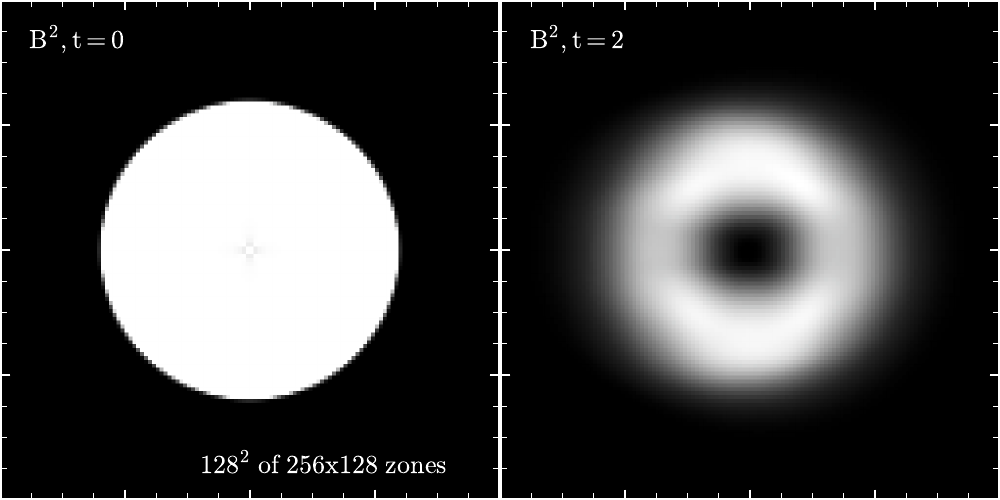}
\includegraphics[height=.3\textheight]{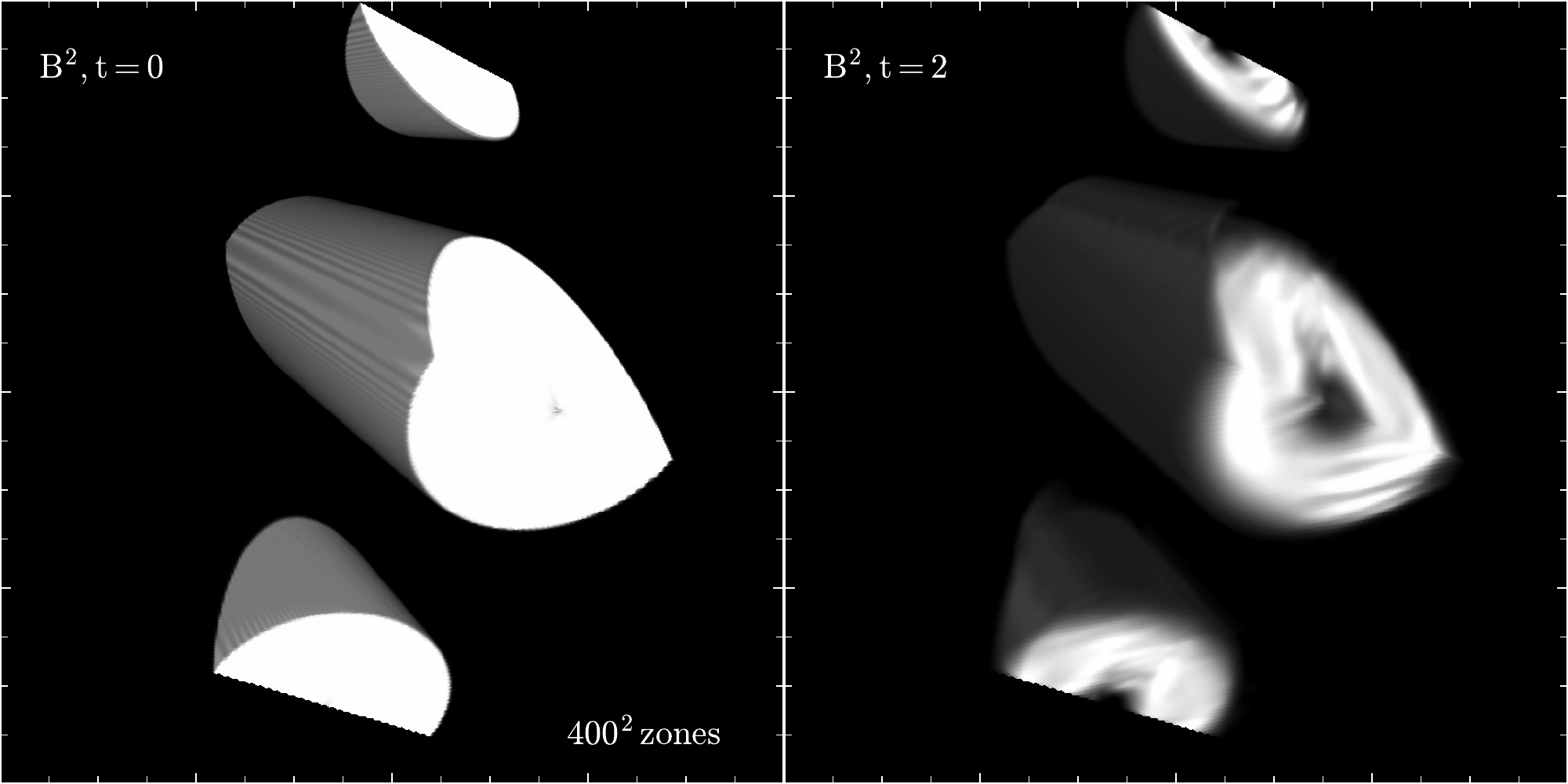}
\caption{Images of magnetic pressure for the advection of a magnetic field loop. The \emph{top left} image shows the initial conditions and the \emph{top right} the solution after two periods across a 2d domain with 256x128 zones. The \emph{bottom left} image shows the initial conditions and the \emph{bottom right} the solution after two periods for a 3d domain with 200$^2$x300 zones.  Each image is scaled linearly from [0,10$^{-6}$].}
\label{fig:Bloop}
\end{figure*}

A powerful test of an MHD code's ability to keep $\nabla \cdot \mathbf{B}$ = 0 is the advection of a weak magnetic field loop.  We use a setup similar to that of GS05 for a 2d calculation.  A periodic box with $L_{X}$ = [-1.,1.] and $L_{Y}$ [-0.5,0.5] over 256x128 zones was initialized with $\rho$ = 1, $P_{gas}$ = 1, $v_{x} = 2$, and $v_{y} = 0.5$.  The magnetic field was derived from a vector potential defined at zone corners as $A_{z} = \text{MAX}(A[R_{0} - r],0)$ where $A = 10^{-3}$ and $R_{0}$ = 0.3. This field produces a line current through the center of the loop and a return current along $R_{0}$, but these features are unresolved on the grid.  Figure \ref{fig:Bloop} shows the 2d result after two periods.

We perform a 3d version of this test, also shown in Figure \ref{fig:Bloop}, following the setup used in GS08.

\subsection{MHD Blast Wave}

\begin{figure*}
\centering
\includegraphics[height=.65\textheight]{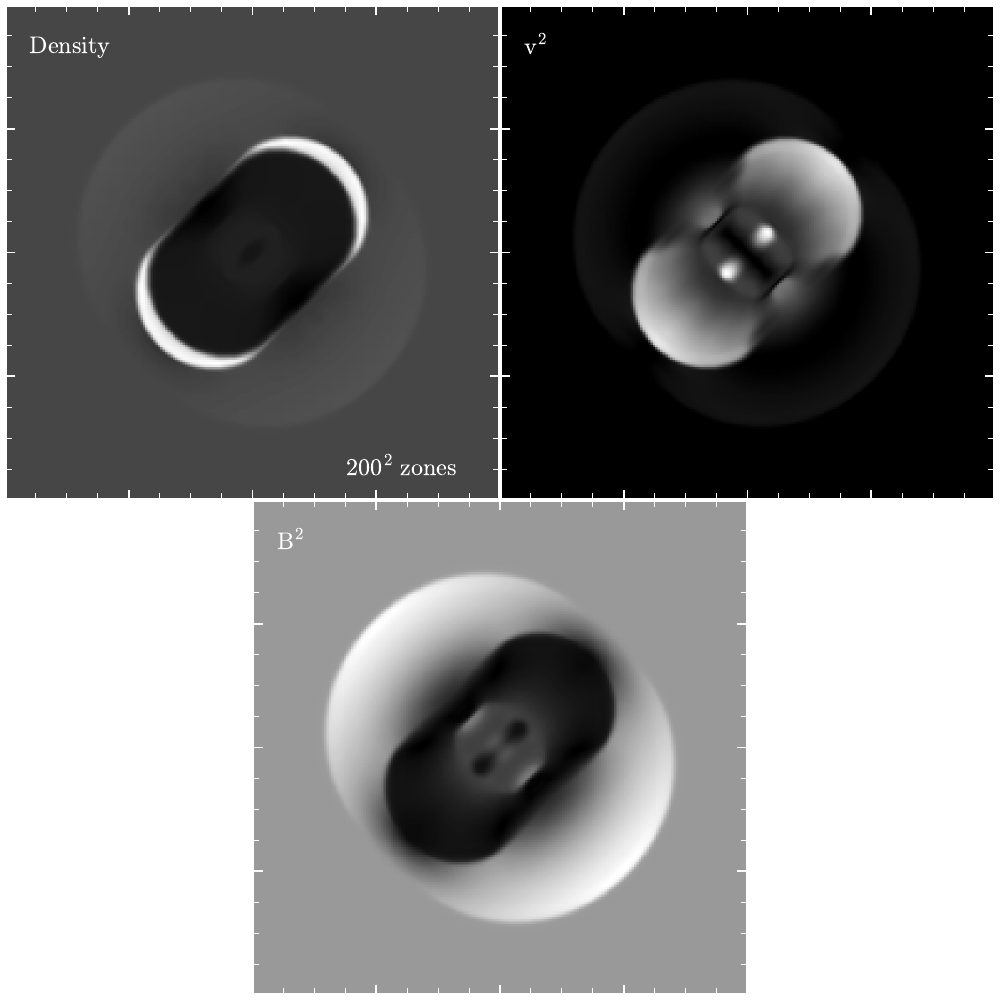}
\caption{Images of selected quantities in a two-dimensional slice at z = 0 for the magnetic blast wave test in three dimensions at t = 0.02. The solution was computed on a domain with 200x300x200 zones.}
\label{fig:3d_blast_wave}
\end{figure*}

We performed a 3d version of the 2d magnetized strong blast wave test as defined in \citet{2000ApJ...530..508L}. The test is performed on a periodic domain with $(L_X, L_Y, L_Z) = (1, 3/2, 1)$ using 200x300x200 zones. The fluid is initialized at rest with $\rho = 1$ and a uniform magnetic field $(B_x, B_y, B_z) = (10/\sqrt{2}, 10/\sqrt{2}, 0)$. The fluid has a pressure $P = 1$, except for in the central region within $r_0 = 0.125$ where $P = 100$. Figure \ref{fig:3d_blast_wave} shows the density, specific kinetic energy, and magnetic energy of the solution in a slice through $z = 0$ at $t = 0.02$.

\subsection{Circularly Polarized Alfv\'{e}n Wave} 

\begin{figure*}
\centering
\includegraphics[height=.23\textheight]{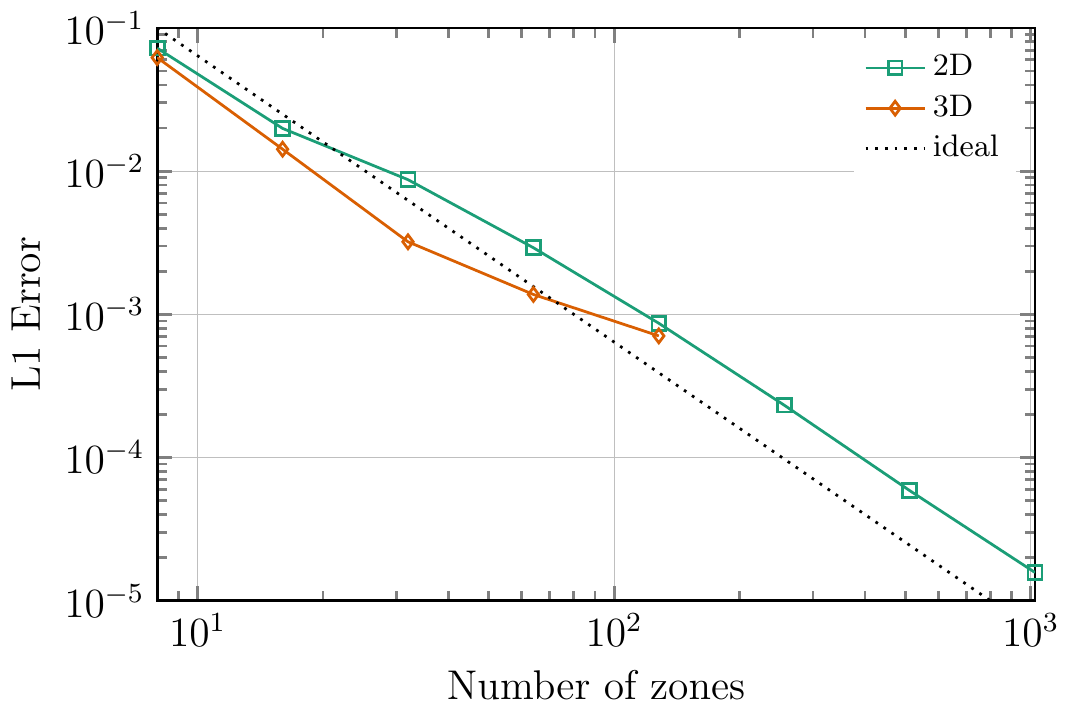}
\includegraphics[height=.23\textheight]{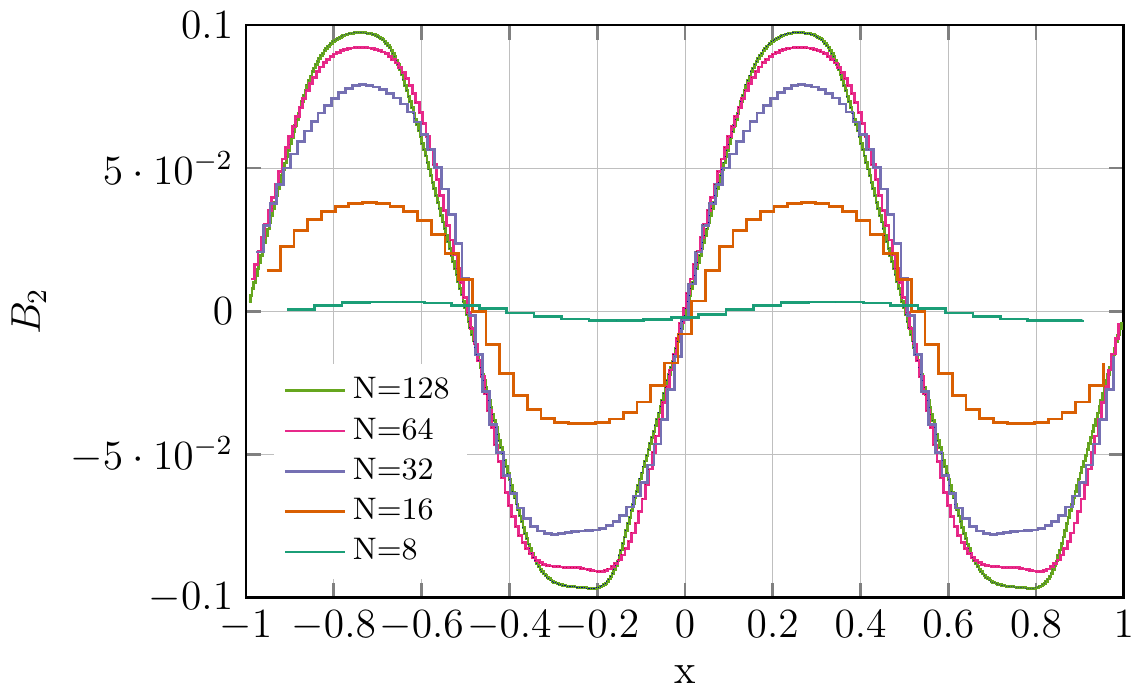}
\caption{Circularly polarized Alfv\'{e}n wave test. \emph{Left}: Convergence of the norm of the $L_1$ error vector for the circularly polarized Alfv\'{e}n wave tests in two and three dimensions. \emph{Right}: Profiles of $B_2$ component of magnetic field for increasing resolutions after propagating five wavelengths across the domain.}
\label{fig:CPA}
\end{figure*}

As a final MHD test, we show the propagation of a circular polarized Alfv\'{e}n wave as described by \citet{2000JCoPh.161..605T}.  This test was used by \citet{2000JCoPh.161..605T} to compare the performance of various approaches to maintaining $\nabla \cdot \mathbf{B}$ = 0.  This test can be done in one or more dimensions, and it can be used for convergence testing as it is an exact nonlinear solution to the equations of MHD.  The grid is initialized with $\rho$ = 1, $P_{gas}$ = 0.1, $v_{y} = 0.1\text{SIN}(2\pi x)$, $B_{y} = 0.1\text{SIN}(2\pi x)$, $v_{z} = B_{z} = 0.1\text{COS}(2\pi x)$, $B_{x} = 1$, and $v_{x}$ = 0.  For the 2d tests we rotate these properties on the grid by an angle of $\theta = \text{TAN}^{-1}(2)$, while in 3d tests we perform the same rotation as 3d tests in \S \ref{wave_conv}.  The grid was a periodic box with $L_{X} =$ [-$\sqrt{5}/2$,$\sqrt{5}/2$] and  $L_{Y} = 0.5*L_{X}$ with 2NxN zones in 2d and $(L_X, L_Y, L_Z) = (3, 3/2, 3/2)$ with 2NxNxN in 3d. The left panel of figure \ref{fig:CPA} shows the convergence of the $L_1$ error vector norm after one wave period for 2d and 3d tests with increasing resolution, where the horizontal axis represents the number of zones across the shorter dimensions. The right panel shows, using every zone in 2d tests, profiles of the in plane transverse component of the magnetic field in the rotated frame ($B_2$) after five wave periods, with the horizontal axis representing the $x$ coordinate in the rotated reference frame. The lack of scatter in these plots demonstrates that the rotated wave fronts remain coherent.


\section{Conclusions}

In this paper we present the design and performance of a new hybrid MPI/OpenMP astrophysical MHD code called WOMBAT.  We are developing WOMBAT for broad application in astrophysics, but especially in support of investigations of cosmological turbulence and the evolution of magnetic fields in galaxy clusters, where conductive fluid behaviors must be captured with good fidelity on a very wide range of scales.  This requirement demands that WOMBAT have exceptional performance and scaling on the latest generation of HPC systems.  We also argue in \S \ref{introduction} that the ability to scale to high thread counts is crucial to maintaining high performance for the target simulations.  This is particularly important for mesh refinement and N-body extensions of WOMBAT currently in development, where load imbalance is unavoidable.  This work will be presented in a follow-up to this paper.  The optimization strategies incorporated into WOMBAT are based on the Patch, a the basic unit of work and domain decomposition within an MPI rank.  Patches are self-contained problems with their own boundary zones and meta-data necessary to evolve them in time. 
These properties make Patches ideal for presenting independent work to threads within a rank.  We presented the SPMD OpenMP design of WOMBAT, where only a single OpenMP parallel region exists for the duration of code execution.  Threads update Patches and perform all boundary communication collaboratively with the Update and MPI-RMA Engines discussed in \S \ref{design_details}.  We present a unique enhancement of the Cray MPICH library through a co-design effort with Cray, Inc. and the University of Minnesota.  The ``thread-hot'' MPI-RMA feature (see \S \ref{mpich_rma}) results in significant speedup of WOMBAT because of its lock-free design.

We show the performance characteristics of WOMBAT on several architectures including the latest generation of Intel Xeon Phi ``Knights Landing'' processors.  WOMBAT scaling on these architectures up to 260K threads on Blue Waters, demonstrates its capabilities and adaptability.

\section{Acknowledgements}

PJM thanks Luiz DeRose (Cray) and John Levesque (Cray) for their support of this project.  JD acknowledges support from the People Programme (Marie Sklodowska Curie Actions) of the European Union’s Eighth Framework Programme H2020 under REA grant agreement no. [658912]. PE is supported by the ITC and Harvard FAS Research Computing.  TWJ and BJO acknowledge support from NSF grant AST1211595.  CN was supported by an NSF Graduate Fellowship under Grant 000039202.  We thank Cray, Inc. for use of their internal systems.  Blue Waters computing resources came through a grant from the Great Lakes Consortium for Petascale Computing. The Blue Waters sustained-petascale computing project is supported by the National Science Foundation (awards OCI-0725070 and ACI-1238993) and the state of Illinois. Blue Waters is a joint effort of the University of Illinois at Urbana-Champaign and its National Center for Supercomputing Applications.


\appendix

\section{MHDTVD} \label{tvd}

Integrating Equation \ref{e:mhd_cons} over a volume element and over a time interval gives 
\begin{align}
	\mathbf{q}_{i}^{n+1} &= \mathbf{q}_{i}^{n} - \frac{\Delta t}{\Delta x}\left(\mathbf{F}_{i+1/2}^{n+1/2} - \mathbf{F}_{i-1/2}^{n+1/2}\right). \label{e:mhd_finite}
\end{align}

In the MHDTVD method, an approximation to $\mathbf{F}_{i+1/2}^{n+1/2}$, referred to as the modified flux $\mathbf{\bar{u}}_{i+1/2}^{n+1/2}$, is computed from
\begin{align}
	\mathbf{\bar{u}}_{i+1/2}^{n+1/2} &= \frac{1}{2}\left(\mathbf{F}(\mathbf{q}_{i}^{n}) + \mathbf{F}(\mathbf{q}_{i+1}^{n}) \right) - \frac{\Delta x}{2\Delta t}
   	\mathbf{f}_{i+1/2}^{n}, \label{e:mhdtvd_flux2} \\
	\mathbf{f}_{i+1/2}^{n} &= \displaystyle\sum_{k=1}^{7}\beta_{k,i+1/2}\mathbf{R}_{k,i+1/2}^{n},  \\
	\beta_{k,i+1/2} &= Q_{k}\left(\frac{\Delta t^{n}}{\Delta x}a_{k,i+1/2}^{n} + \gamma_{k,i+1/2}\right)\alpha_{k,i+1/2} - (g_{k,i} + g_{k,i+1}), \\
	\alpha_{k,i+1/2} &= \mathbf{L}_{k,i+1/2}^{n} \cdot (\mathbf{q}_{i+1}^{n} - \mathbf{q}_{i}^{n}), \\
	\gamma_{k,i+1/2} &= \begin{cases}
    					\frac{g_{k,i+1} - g_{k,i}}{\alpha_{k,i+1/2}} & \text{for } \alpha_{k,i+1/2} \ne 0, \\
                         0 & \text{for } \alpha_{k,i+1/2} = 0
        			    \end{cases}, \\
	g_{k,i} &= \text{SIGN}(\tilde{g}_{k,i+1/2})\text{ SWEBY }_{\text{limiter}}\left(\tilde{g}_{k,i+1/2},\tilde{g}_{k,i-1/2}\right), \\
\tilde{g}_{k,i+1/2} &= \frac{1}{2}\left[Q_{k}\left(\frac{\Delta t^{n}}{\Delta x}a_{k,i+1/2}^{n}\right) - \left(\frac{\Delta t^{n}}{\Delta x}a_{k,i+1/2}^{n}\right)^{2}\right]\alpha_{k,i+1/2}, \\
	Q_{k}(\chi) &= \begin{cases}
                            \frac{\chi^{2}}{4\epsilon_{k}} + \epsilon_{k} & \text{for } |\chi| < 2\epsilon_{k}, \\
                            |\chi| & \text{for } |\chi| \geq 2\epsilon_{k}
               \end{cases}.
\end{align}
The right-handed eigenvector, $\mathbf{R}_{k,i+1/2}^{n}$, and characteristics, $\alpha_{k,i+1/2}$, are from \citet{Cargo:1997:RMI:269910.269926}.  The primitive variables at zone interfaces, used to construct $\mathbf{R}_{k,i+1/2}^{n}$ and $\alpha_{k,i+1/2}$, come from the averaging scheme also described in \citet{Cargo:1997:RMI:269910.269926}.  The purpose of $\epsilon_{k}$ is to add a controlled amount of dissipation into each wave to ensure that $Q_{k}(\chi)$, referred to as the coefficient of numerical viscosity, is continuous and positive \citep{1998CNSNS...3...82Z}.  This eliminates spurious oscillations that can occur when there is an entropy violation across a discontinuity.  The value of $\epsilon_{k}$ must satisfy $0 \le \epsilon_{k} < 0.5$, and the optimal value depends on the number of dimensions and complexity of flows in the calculation.

Under certain circumstances, Roe-type methods like MHDTVD will produce unphysical densities or pressures \citep{1991JCoPh..92..273E}.  A typical solution to this problem is to define floor values for density and pressure that are applied when exceeded.   WOMBAT uses this approach, but additionally offers a set of user-defined floor values, called the protection floor, that will automatically switch to another Riemann solver that does not have this issue.  Similar to the approach of GS08, we substitute the MHDTVD fluxes with the more diffusive HLL fluxes \citep{1991JCoPh..92..273E} under the rare conditions when the protection floor is exceeded.  The modified flux $\mathbf{\bar{u}}_{i+1/2}^{n+1/2}$ is computed for the HLL scheme as 
\begin{align}
	\mathbf{\bar{u}}_{i+1/2}^{n+1/2} &= \frac{b^{+}\mathbf{F}(\mathbf{q}_{i}^{n}) + b^{-}\mathbf{F}(\mathbf{q}_{i+1}^{n})}{b^{+} - b^{-}} +
                                  \frac{b^{+}b^{-}}{b^{+} - b^{-}}(\mathbf{q}_{i+1}^{n} - \mathbf{q}_{i}^{n}), \\
	b^{+} &= \text{MAX}\{\text{MAX}(a_{max},v_{x,i+1}^{n} + c_{f,i+1}^{n}),0\}, \\
	b^{-} &= \text{MIN}\{\text{MIN}(a_{min},v_{x,i-1}^{n} - c_{f,i11}^{n}),0\},
\end{align}
where $a_{max}$ and $a_{min}$ are the maximum and minimum eigenvalues.  Note that the HLL fluxes do not rely on an eigensolution to the MHD equations, which makes them more diffusive than the MHDTVD fluxes.  Consequently, we apply them as infrequently as possible; so only to avoid unphysical behaviors.

\bibliographystyle{apj} \bibliography{wombat}

\end{document}